\documentclass[pra,aps,amsmath,amssymb,amsfonts,twocolumn,nofootinbib,longbibliography]{revtex4}
\usepackage{}
\usepackage{amssymb}
\usepackage{bm,mathrsfs}
\usepackage{graphicx}
\usepackage{epsfig}
\usepackage{amsmath,bbm}
\usepackage{amsfonts,amssymb}
\usepackage{times}
\usepackage{verbatim}
\usepackage[sort&compress]{natbib}
\usepackage{amsmath}
\usepackage{bm}
\usepackage{float}
\usepackage{textgreek}
\allowdisplaybreaks[4]
\usepackage[colorlinks,breaklinks,linkcolor=blue,anchorcolor=blue,citecolor=blue,urlcolor=magenta]{hyperref}

\usepackage{color}
\definecolor{zzz}{rgb}{0.9,0.0,0.4}

\begin{document}
\title{Amplification of Weak Forces via Parametric Interactions and Non-Markovian Effects in Cavity Optomechanics}

\author{Y. F. Li,$^{1}$ Ze Wang,$^{1}$ W. Y. Hu,$^{1}$ Yan-Hui Zhou,$^{2,}$\footnote{\textcolor{zzz}{Corresponding author: yanhuizhou@126.com }} Cheng Shang,$^{3,}$\footnote{\textcolor{zzz}{Corresponding author: cheng.shang@riken.jp}}  and H. Z. Shen$^{1,}$\footnote{\textcolor{zzz}{Corresponding author: shenhz458@nenu.edu.cn }}}
\affiliation{$^1$Center for Quantum Sciences and School of Physics, Northeast Normal University, Changchun 130024, China\\
$^2$Quantum Information Research Center and Jiangxi Province Key Laboratory of Applied Optical Technology, Shangrao Normal University, Shangrao 334001, China\\
$^3$Analytical quantum complexity RIKEN Hakubi Research Team,
RIKEN Center for Quantum Computing (RQC), Wako, Saitama 351-0198, Japan}
\date{\today}

\begin{abstract}
Weak force amplification describes the process of amplifying a faint low-frequency signal by means of an additional high-frequency modulation, which plays a vital role in quantum sensing and high-precision measurement. However, the potential enhancement of weak-force amplification in non-Markovian environments has received little attention. In this paper, we firstly study the amplification of weak forces within cavity-optomechanical systems incorporating a degenerate optical parametric amplifier (DOPA) under the Markovian assumption. The results show that the weak force can be effectively amplified by using two high-frequency signals via vibrational resonance through adjusting the strength and phase of the DOPA with different pumping frequencies. Moreover, we extend the study of the amplification of the weak force to the non-Markovian environment, which consists of a collection of infinite oscillators. We illustrate that the amplification exhibits a conversion from the non-Markovian regime to Markovian regime by controlling environmental spectral width. This conversion leads to enhancements of amplification, which originates from the excitation backflow obtained through the interaction between the cavity and non-Markovian environment. By controlling DOPA to amplify weak forces, the study achieves amplification in the non-Markovian regime, offering new directions for quantum optics research.
\end{abstract}

\maketitle
\section{Introduction}
The growing interest in weak signal amplification in recent years is driven by its broad potential for application across numerous scientific disciplines. Vibrational resonance (VR) is an efficient nonlinear amplification mechanism \cite{Landa33433,Gitterman34355,Chizhevsky91220602,Chizhevsky73022103,Yao83061122}. It is found that the response of the system to a weak low-frequency signal can become maximal by altering the input amplitudes at a high-frequency periodic signal. Due to the fact that the high-frequency perturbation in VR is deterministic and easy to control in experiments, and bichromatically driven systems are prevalent in nature and research fields, VR has attracted continuous research attention. In 2000, Landa and McClintock first identified VR in bistable systems \cite{Landa33433}. Theoretical approaches have been developed to study VR \cite{Chizhevsky181767,Zaikin66011106,Blekhman39421,Shan109023511}. Furthermore, VR is demonstrated experimentally in various systems \cite{Chizhevsky91220602,Baltans67066119,Chizhevsky70062101,Ullner312348,Chowdhury112400,Chizhevsky73022103,Chizhevsky90042924,Chizhevsky92032902}.

By coupling mechanical and optical modes through radiation pressure, cavity-optomechanical systems have developed into a highly attractive research area over the past decades. Many fields have illustrated its advantages, including quantum precision measurements \cite{Zhao63224211,Zhang86053806,Gavartin7509,Schliesser5509,Esfahani14085017,Krause6768,Liu2119555,Eichenfield459550}, quantum computation or communication \cite{Stannigel105220501,Naeini13013017,Zhang48015502}, and even fundamental tests of quantum mechanics \cite{Kleckner10095020,Chen46104001,Pikovski8393,Isart107020405}. Due to the optomechanical nonlinearities, they are excellent platforms for studying various nonlinear phenomena. Considering the capability of optomechanical systems to engage with a multitude of signals, containing both mechanical and optical varieties, it is fascinating to explore VR within these systems.

Cavity optomechanics \cite{Shang25882,Shange00606} research has proposed numerous schemes to enhance the performance of the cavity-optomechanical systems by using nonlinear optical interactions, such as parametric amplification and optical Kerr effect \cite{Xiong58050302,Wang17024009,Bartolo94033841,Zhou421289,Ilchenko92043903,Jiao97013843,Li103053522,Huang121153601,Wang98035108,Xu90043822,zx1,zx2,zx3,Sun043715}. An optical parametric amplifier (OPA) inside the optomechanical cavity (pumped by an external laser) can directly lead to optical amplification and modulate the optomechanical coupling in a way analogous to periodic cavity driving \cite{Adamyan92053818,Hu100043824,Lu114093602,Savelev85013811,Tholen137014019,Liu53624000182024,Yang023703,Luan2350021,Shen2350029,Zhou064009,Liu2300422}. The OPA can generate down-converted photons with nearly ideal single or dual-mode squeezing. The degenerate case (corresponding to DOPA) can manipulate dynamical instabilities and nonlinear dynamics in the system, enabling better control and optimization in advanced physical applications \cite{Xuereb86013809,Qin120093601}. However, there is a lack of theoretical work on the weak force amplification with parametric interactions.

For open quantum systems \cite{breuer2002,Weiss2008,Shen052122}, the Markovian approximation is valid only under the condition that the coupling between the system and its environment \cite{Cui128314} is weak, where the characteristic time of the environment is much smaller than that of the quantum system under consideration. In specific cases like two-state systems, harmonic oscillators, coupled cavities \cite{Leggett5911987,breuer1032104012009,Tan032102,Cavaliere8762025,Wilkey63082023,Yang72822024,Yang852025,Xin105053706,shen960338052017,rivas1050504032010,wolf1011504022008,chruscinski1121204042014,Triana116183602,Cheng623678,shen990321012019,Zhang063853,shen97042121,shen98062106,Ding111814,Sinha124043603,Wu103010601,Mu94012334,Li109023712,Shen98023856,Shen101013826,Shen1070537052023,shen1050237072022,Zhang1090337012024,Yang1090537122024,tang1100437062024},  the Markovian approximation may be restricted, necessitating the consideration of the influences of non-Markovian effects \cite{Luan2503,Zhang033701,Sun063713,Cui032209,Sun063704,Shen043714,Cui042129,Li56142022,Shen315042019,Shen28522018,Shen042129,ShenNHgain2025,Shen012107,Shene71535,Wang34001} on the system dynamics. Moreover, we demonstrate that non-Markovian processes are beneficial for quantum information processing, including state engineering, control, and enhancing channel capacity \cite{caruso8612032014,darrigo3502112014,lofranco900543042014,bylicka457202014,xue860523042012}, with experimental realization confirmed \cite{hoeppe1080436032012,madsen1062336012011,Guo2021126,Li2022129,Xiong2019100,Xu201082,Uriri2020101,Liu2020102,Anderson199347,breuer880210022016,Vega015001}.

In this paper, we investigate the amplification of weak forces in cavity-optomechanical systems through parametric interactions and non-Markovian effects. Specifically, we examine how a DOPA influences signal amplification by applying two forms of high-frequency modulation: mechanical signals and detuned optical signals. The results show that the amplification induced by two types of signals can be effectively enhanced by controlling the strength and phase of the DOPA. With the amplification of a weak mechanical signal by a high-frequency mechanical signal, a dramatic improvement in the relative response amplitude ($A/A_{0}$) generated by VR can be clearly observed in the presence of the DOPA. In this process, we use $A/A_{0}$ to denote a measure of the amplitude of the signal frequency component in the system response. A weak mechanical signal is amplified by an optical signal, where $A/A_{0}$ can be further amplified by modulating the pump frequency. We also compare the differences in $A/A_{0}$ under various pump frequencies.
Moreover, the relative response amplitude $A/A_{0}$ is generalized to the non-Markovian regimes and compared with Markovian approximation in the wideband limit. In the non-Markovian environment, $A/A_{0}$ decreases when the weak mechanical signal is amplified by a high-frequency mechanical signal. When the weak mechanical signal is amplified by an optical signal, $A/A_{0}$ shows variations under different pumping frequencies, which exhibit as either an increase or a decrease.

The rest of this paper is organized as follows. In Sec. II, the physical model is introduced and some results are given under the Markovian approximation. We discuss the influences of the DOPA on the amplification of a weak low-frequency mechanical signal by two types of high-frequency signals. In Sec. III, we extend the weak-force amplification system under the Markovian approximation to the non-Markovian one and compare it with that in the Markovian regime (the spectral width approaching infinity). In Sec. IV is devoted to the experimental Implementation. The conclusions and discussions are shown in Sec. IV.

\section{Weak-force Amplification with DOPA under the Markovian approximation in Cavity Optomechanical System}
\subsection{Influences of the DOPA on the amplification of a weak low-frequency mechanical signal by a high-frequency mechanical signal}

\begin{figure}[h]
\centerline{
\includegraphics[width=8cm, height=4.4cm, clip]{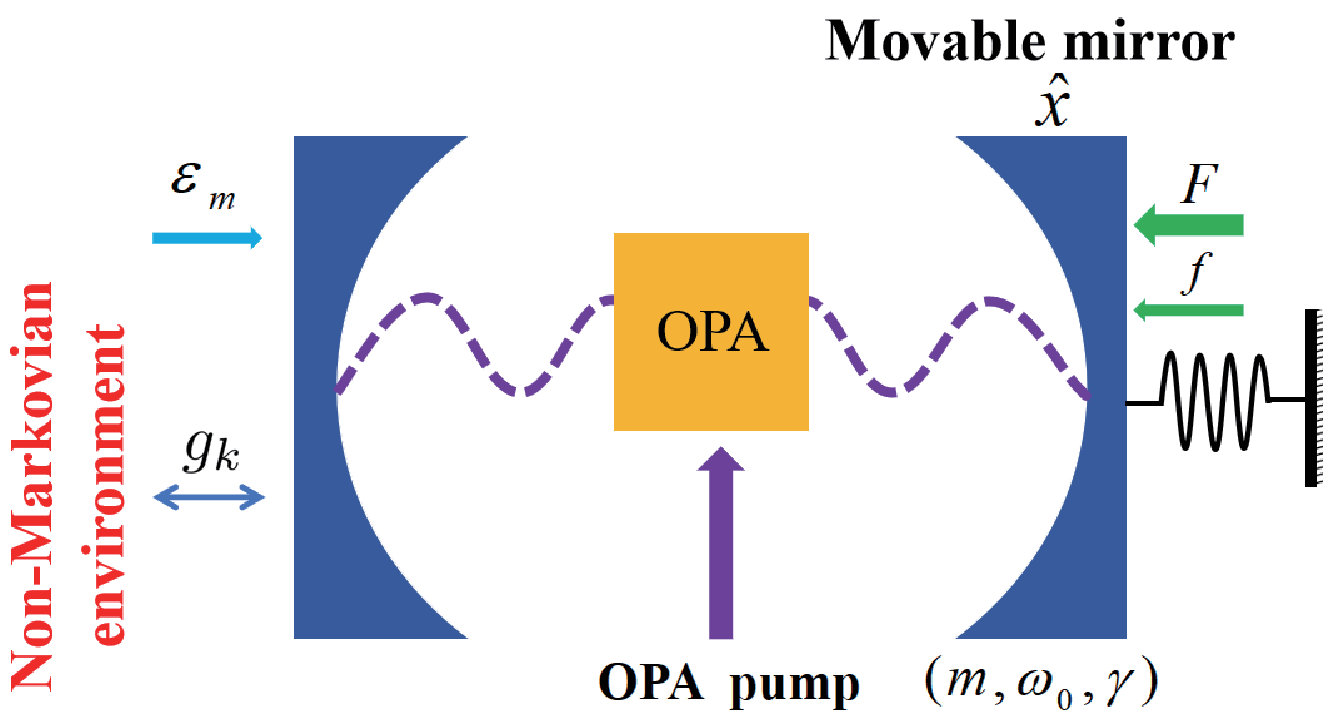}}
\caption{Setup of the optomechanical system driven by four fields. A cavity-optomechanical system is driven by an optical field (amplitude $\varepsilon _m$ and frequency $\omega _m$) with the DOPA (coupling coefficient $G{e^{i\theta }}$ and frequency $\omega _g$) inside the cavity and two mechanical signals (amplitudes $f$ and $F$). The dissipation induced by the coupling between the cavity and external environment can be Markovian or non-Markovian with the coupling strength ${{g_k}}$, which depends on the specific structure of the environment.}\label{model}
\end{figure}

\begin{figure*}[t]
\centerline{
\includegraphics[width=17.8cm, height=6.4cm, clip]{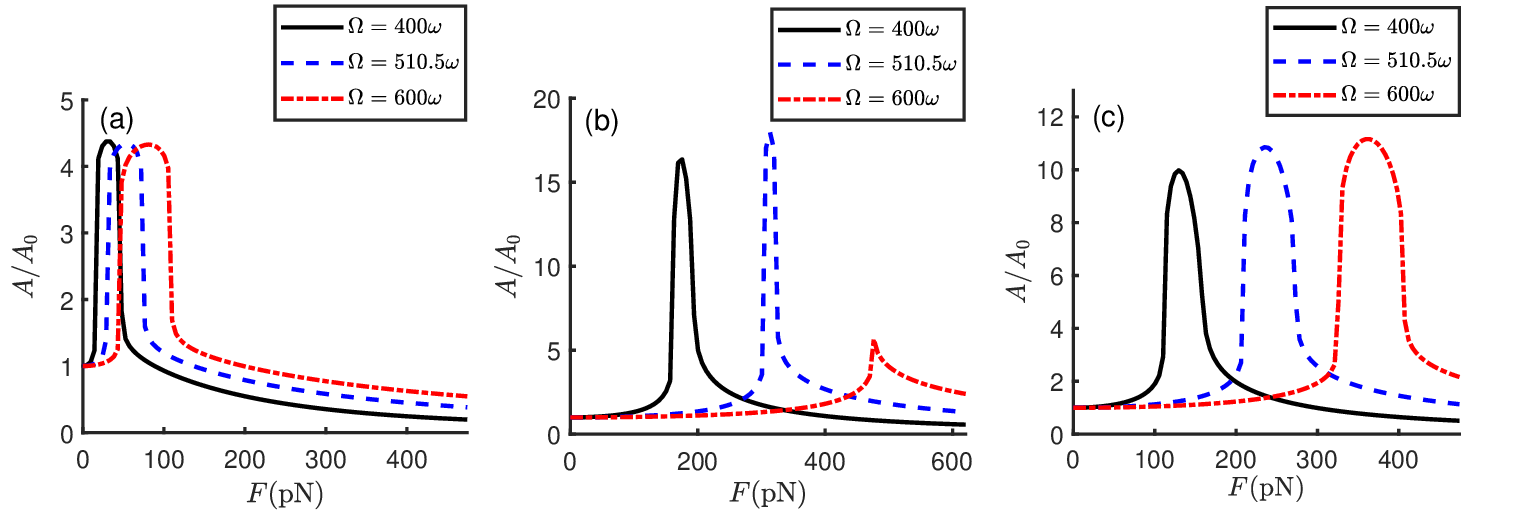}}
\caption{The influences of the modulation frequency $\Omega$ on the amplification ($\omega _g=2{\omega _m}$). The relative response amplitude $A/A_{0}$ of the system is determined by Eq.~(\ref{q}) as a function of $F$, where the strength and phase of the probe field of the DOPA are fixed as (a) $G = 0,\theta = 0$; (b) $G = 0.02 \omega_0$, $\theta = 0$; (c) $G = 0.02 \omega_0$, $\theta = 3\pi /2$. Other parameters are $\omega_{0}=2 \pi \times 7.9 \mathrm{MHz}$, $\kappa=6.2 \omega_{0}$, $\xi_0=2 \pi \times 0.63 \mathrm{MHz}$, $\Delta_0=5.58 \omega_{0}$, $P_{m}=3.9 \mathrm{nW}$, $\gamma=0.5 \omega_{0}$, $f=0.22 \mathrm{pN}$, and $\omega=$ $0.01 \omega_{0}$.}\label{sgzfig2}
\end{figure*}

As shown in Fig.~\ref{model}, we consider a cavity-optomechanical system respectively driven by multiple signals: an optical driving field with amplitude $\varepsilon _m$ and the frequency $\omega _m$, the DOPA inside the cavity with the frequency $\omega _g$ \cite{Adamyan92053818,Hu100043824,Lu114093602,Savelev85013811,Tholen137014019,Liu53624000182024}, and two mechanical forces with amplitudes $f$ and $F$. In our system, the mechanical oscillator is the condition for VR. The resonance frequency of this mechanical oscillator is $\omega_{0}$, where the mechanical motion can couple with the optical cavity via radiation pressure. This coupling endows the movable mirror with nonlinear characteristics. VR emerges in our system when a high-frequency perturbation is applied to the nonlinear system with its intensity matching both the properties of the system and the weak signal to be amplified. In this case, the response of system to the weak signal can be enhanced, thereby producing an amplified output at the signal frequency \cite{yang106712024}. Specifically, the mechanical force with a lower frequency $\omega$ and amplitude $f$ represents the weak periodic signal intended for amplification. Conversely, the mechanical force with a higher frequency $\Omega$ and amplitude $F$ serves as the modulation signal, enabling the realization of the VR effect. For the VR effect to be obtained, we assume $\Omega \gg \omega$. The DOPA inside the cavity is excited by a pump field with the frequency $\omega _g$. The cavity dissipation occurs in two ways: the Markovian and non-Markovian processes. To be specific, we first focus on the Markovian case, while subsequently investigating the influences of the non-Markovian effects on the weak-force amplification.

In the rotating frame at the frequency ${{\omega _m}}$, we now begin with the system's total Hamiltonian (setting $\hbar  \equiv 1$)

\vspace{-1\baselineskip}
\begin{align}
\hat H =& \frac{1}{2}m\omega _0^2{{\hat x}^2} + \frac{{{{\hat p}^2}}}{{2m}} + \hbar \Delta_0 {{\hat a}^\dag }\hat a - \hbar \xi{{\hat a}^\dag }\hat a\hat x \nonumber\\
&- i\hbar \sqrt \kappa  {\varepsilon _m}( {\hat a - {{\hat a}^\dag }} ) + {f}\cos ( {{\omega}t} )\hat x + {F}\cos ( {{\Omega}t} )\hat x\nonumber\\
&+ i\hbar G({{\hat a}^{\dag 2}}{e^{i\theta }} - H.c.),\label{H1}
\end{align}where $\hat{x}$ and $\hat{p}$ respectively denote the position and momentum operators of the mechanical oscillator with commutation relations $[ {\hat x,{\kern 1pt} \hat p} ] = i\hbar$ \cite{Davuluri11264002}. Here, $\hat a( {{{\hat a}^\dag }} )$ is the annihilation (creation) operator of the optical cavity field with resonance frequency ${{\omega _a}}$. The mechanical oscillator with mass $m$ vibrates with the frequency $\omega_{0} /(2 \pi)$. ${\varepsilon _m} = \sqrt {2{P_m}/ \hbar{\omega _a}}$ is the amplitude of the driving field. $\xi  = {\omega _a}/L$ denotes the optomechanical coupling constant ($P_{m}$ is the power of the driving field, while $L$ represents the characteristic length of the optical cavity). Herein, $\Delta_0=\omega_{a}-\omega_{m}$ is the frequency detuning between the resonant frequency of the optical cavity and driving field. The last term describes the coupling of the intracavity field with the DOPA (pump frequency $\omega_g$). $G$ is the strength of the DOPA, which is proportional to the pump power driving amplitude. $\theta $ denotes the phase of the DOPA \cite{Shahidani88053813}. We have assumed that this DOPA is excited by a pump field with the frequency $\omega _g= 2{\omega _m}$.
The Heisenberg-Langevin equations of the system operators read

\vspace{-1\baselineskip}
\begin{small}
\begin{align}
\frac{{d\hat a}}{{dt}} &= - ( {i\Delta_0  + \frac{\kappa }{2}} )\hat a + i\xi \hat a\hat x+ 2G{e^{i\theta }}{{\hat a}^\dag } + \sqrt \kappa  ( {{\varepsilon _m} + {{\hat a}_{{\rm{in}}}}} ),\label{Heq1}\\
\frac{{d\hat x}}{{dt}} &= \frac{{\hat p}}{m},\label{Heq2}\\
\frac{{d\hat p}}{{dt}} &=  - {\gamma}\hat p - m\omega _0^2\hat x + \hbar\xi{{\hat a}^\dag }\hat a- {f}\cos ( {{\omega}t} ) - {F}\cos ( {{\Omega}t} ),\label{Heq3}
\end{align}
\end{small}where $\hat{a}_{\text {in }}$ is the quantum fluctuation part of the optical driving field. $\gamma$ denotes the mechanical damping rate. $\kappa$ represents the optical decay rate under the Markovian approximation.
Since our system is strong driving and weak coupling between the optical mode and mechanical mode, we neglect quantum fluctuations of optics and mechanics, and write the classical Langevin equations by replacing quantum operators in Eqs.~(\ref{Heq1})-(\ref{Heq3}) with classical complex variables $\hat{a} \rightarrow \beta$, $\hat{x} \rightarrow x$, and $\hat{p} \rightarrow p$ \cite{Vitali98030405,Aspelmeyer861391}. Thus, we arrive at the equations of motion for the mean-field amplitudes of the system operators

\begin{small}
\begin{align}
\dot \beta &=  - ( {i\Delta_0  + \frac{\kappa }{2}} )\beta  + i\xi \beta x + \sqrt \kappa  {\varepsilon _m} + 2G{e^{i\theta }}{\beta ^*}
,\label{Heq4}\\
\dot x &= \frac{p}{m}
,\label{Heq5}\\
\dot p &=  - {\gamma}p - m\omega _0^2x +\hbar \xi|\beta {|^2}- {f}\cos ( {{\omega}t} ) - {F}\cos ( {{\Omega}t} )
.\label{Heq6}
\end{align}
\end{small}

In this work, we focus on the system response to the mechanical signal $f$, and approximately derive the equation of motion merely for the mechanical mode under $\kappa  \gg$ (${\gamma}$, $\xi_0$), where the dynamics of the optical mode is much faster than that of the mechanical mode. In this case, the steady value of $\beta$ is derived as
\begin{equation}
\beta  = \frac{{2\sqrt \kappa  {\varepsilon _m}(4{e^{i\theta }}G - 2ia + \kappa )}}{b}
,\label{alphastable}
\end{equation}
where $a = {\Delta _0} - \xi x$ and $b = 4{a^2} - 16{G^2} + {\kappa ^2}$. We can adiabatically eliminate the optical mode by substituting Eq.~(\ref{alphastable}) into the equations of motion for the mechanical mode in Eq.~(\ref{Heq6}), and then write the equation of motion merely for the mechanical mode as
\begin{small}
\begin{equation}
\begin{aligned}
\ddot x + \gamma \dot x \ =&  - \omega _0^2x - \frac{{f\cos ( {\omega t} )}}{m} - \frac{{F\cos ( {\Omega t} )}}{m}\\
&+ \frac{{c( { - 2ia + \kappa  + 4G{e^{i\theta }}} )( {2ia + \kappa  + 4G{e^{ - i\theta }}} )}}{{{b^2}m}}
,\label{x1markovian}
\end{aligned}
\end{equation}
\end{small}where $c = 4\hbar\varepsilon _m^2\xi  \kappa$. With this, we separate the slow and fast motions by \cite{Gitterman34355,Ghosh88042904}

\vspace{-1\baselineskip}
\begin{equation}
\begin{aligned}
x(t)=X(t)+\varphi(t, \tau=\Omega t)
,\label{xfastslow}
\end{aligned}
\end{equation}
where $X(t)$ is the slow motion component of the system response caused by $f$, while $\varphi(t,\tau)$ denotes the fast motion caused by $F$. The mean value of $\varphi(t,\tau)$ with respect to $\tau$ is given by

\vspace{-1\baselineskip}
\begin{equation}
\begin{aligned}
\langle \varphi \rangle  = \frac{1}{{2\pi }}\int_0^{2\pi } \varphi  ( {t,\tau  = {\Omega}t} )d\tau  = 0
.\label{fai1}
\end{aligned}
\end{equation}

\begin{figure*}[t]
\centerline{
\includegraphics[width=17.5cm, height=6cm, clip]{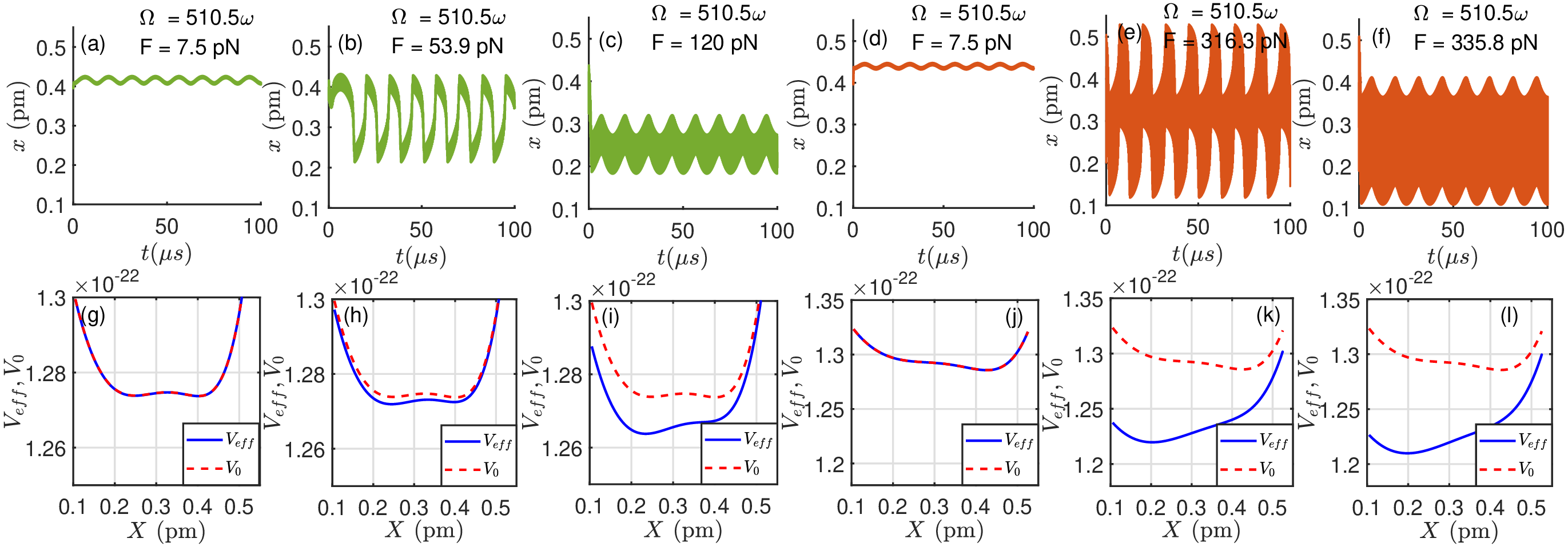}}
\caption{The figure shows the mechanical position $x$ in Eqs.~(\ref{Heq4})-(\ref{Heq6}) as a function of the time $t$ at different high-frequency force $F$ for $\Omega=510.5 \omega$, where the strength and phase of the DOPA take (a)-(c) $G = 0$, $\theta = 0$; (d)-(f) $G = 0.02 \omega_0$, $\theta = 0$; (g)-(l) The effective potential functions $V_{\text {eff }}(X)$ corresponding to Fig.~\ref{g0g0.02xu}(a)-(f) are obtained according to Eq.~(\ref{ueff1}) ($\omega _g=2{\omega _m}$). $V_{0}(X)$ in Eq.~(\ref{u0}) is the potential function in the absence of $F$. Other parameters are the same as  Fig.~\ref{sgzfig2}.}\label{g0g0.02xu}
\end{figure*}

To quantify the response of the system to the weak signal, we define \cite{Yao83061122,Ghosh88042904}

\vspace{-1\baselineskip}
\begin{equation}
\begin{aligned}
A = \frac{{\sqrt {f_s^2 + f_c^2} }}{{{f}}}
,\label{q}
\end{aligned}
\end{equation}
which denotes the ratio of the Fourier coefficient of the system response at the frequency of the weak signal to be amplified and the signal amplitude. $f_{s}$ and $f_{c}$ respectively represent the sine and cosine Fourier components of the system response $x(t)$ at the signal frequency $\omega$, which are shown as follows

\vspace{-1\baselineskip}
\begin{align}
{f_s} = \frac{2}{{nT}}\int_0^{nT} x (t)\sin ( {{\omega}t} )dt
,\label{fs}\\
{f_c} = \frac{2}{{nT}}\int_0^{nT} x (t)\cos ( {{\omega}t} )dt
,\label{fc}
\end{align}
where $T = {{2\pi } \mathord{\left/
 {\vphantom {{2\pi } \omega }} \right.
 \kern-\nulldelimiterspace} \omega }$. In our model, the nonlinear optomechanical interaction enables the system to give an amplified response to an injected mechanical signal, even in the absence of VR. To quantify the amplification effect of the mechanical signal purely by VR, we define a relative response amplitude $A/A_{0}$, where $A_{0}$ represents the amplitude response of the system in the absence of the high-frequency signal $F$.

In our numerical simulations, we have chosen the parameters as $\omega_{0}=2 \pi \times 7.9 \mathrm{MHz}$, $m=43 \mathrm{pg}$, $\gamma=2 \pi \times 3.95 \mathrm{MHz}$, $\omega_{a}=2 \pi \times 194 \mathrm{THz}$, and $\kappa=2 \pi \times 50 \mathrm{MHz}$. Additionally, $\xi  = 2\pi  \times 90.3{\rm{GHz}}/{\rm{nm}}$ corresponds to single-photon optomechanical coupling coefficient ${\xi _0}=\xi \sqrt { \hbar/m{\omega _0}}=2 \pi \times 0.63 \mathrm{MHz}$. The baseline values of \(\omega_0\), \(m\), \(\gamma\), \(\omega_a\), \(\kappa\), and \(\xi\) are the same, up to notation changes, as those used in the earlier optomechanical VR study of Ref.~\cite{Shan109023511}. The strong dispersive coupling scale is further motivated by photonic-crystal optomechanical cavities, where the coupling coefficient can reach the order of \(10^2~\mathrm{GHz/nm}\). The optical decay rate used in the simulations, \(\kappa/2\pi=48.98~\mathrm{MHz}\), is chosen for a high-\(Q\) silicon nitride material optical cavity and should not be identified with the specific low-\(Q\) zipper-cavity linewidth \(\Gamma/2\pi\sim6~\mathrm{GHz}\) reported for the particular device studied in Ref.~\cite{Eichenfield459550}. Specifically, in their dynamical analysis of a slot-type photonic crystal optomechanical cavity with dispersive coupling as large as 940 GHz/nm, Li et al. explicitly employed an optical decay rate of $\kappa/2\pi = 387$ MHz in Ref.~\cite{li23844182010}. Thus, the parameter set used here is an experimentally motivated high-\(Q\) optomechanical regime rather than a direct reproduction of one particular experimental device.

To see the influences of the DOPA on the enhancement of weak signal, the relative response amplitude $A/A_{0}$ of the system to the weak signal $f$ is investigated as a function of the amplitude of the high-frequency force $F$ at a few different frequencies ($\Omega = 400 \omega, 510.5 \omega$, and $600 \omega$) shown in Fig.~\ref{sgzfig2}. In Fig.~\ref{sgzfig2}(a), we discuss that the relative response amplitude $A/A_{0}$ varies with the high-frequency force $F$ without the participation of the DOPA, i.e., the strength $G = 0$ and phase $\theta  = 0$. Under the resonator stationary, all three curves exhibit resonance peaks, indicating the occurrence of the VR phenomenon. The resonance peak shifts towards larger values of $F$ as the modulation frequency $\Omega$ increases. Obviously, the weak force can be dramatically amplified by adding a high-frequency mechanical force. As shown in Fig.~\ref{sgzfig2}(b), we present the changes in the relative response amplitude $A/A_{0}$ in the presence of the DOPA. Specifically, when $G = 0.02 \omega_0$, $\theta  = 0$, and $\Omega = 400 \omega$, $A/A_{0}$ can increase from 4.4 to 16.4. When the frequency of high-frequency mechanical signal $\Omega = 510.5 \omega$, $A/A_{0}$ also can increase from 4.3 to 18. But at the higher frequency of $\Omega = 600 \omega$, the relative response amplitude $A/A_{0}$ exhibits a decreasing trend. Moreover, figure~\ref{sgzfig2}(c) shows that the relative response amplitude $A/A_{0}$ can also be tuned by controlling $\theta $. When $\theta $ changes from $\theta  = 0$ to $\theta  = 3\pi /2$, in the case of $G = 0.02 \omega_0$ and $\Omega = 600 \omega$, the maximum value of $A/A_{0}$ increases to 11. Whereas at $\Omega = 400 \omega$ and $510.5 \omega$, although the maximum value decreases slightly, the need for values of $F$ decreases. This feature indicates that, under the action of the DOPA, the weak signal can be stably enhanced. We see that the relative response amplitude of the system to the weak signal is sensitive to the variations of the strength and phase of the DOPA, which indicates the advantage of applying a hybrid nonlinear system.

To deeply study the physical reason behind $A/A_{0}-F$ curves in the presence of the DOPA, we respectively show the dynamics $x(t)$ at three points on $A/A_{0}-F$ curves for $\Omega = 510.5 \omega$, $G = 0$, $\theta =0$ in Fig.~\ref{g0g0.02xu}(a)-(c), while $G = 0.02 \omega_0$, $\theta = 0$ in Fig.~\ref{g0g0.02xu}(d)-(f). Moreover, we plot the corresponding effective potential function $V_{\text {eff }}(X)$ in Fig.~\ref{g0g0.02xu}(g)-(l). When the amplitude of the high-frequency force is low (e.g., $F=7.5 \mathrm{pN}$), the position of the mechanical system oscillates with a small amplitude around a stable state, which corresponds to the right well of the potential function in Fig.~\ref{g0g0.02xu}(g) and (j). When the amplitude rises to $F=53.9 \mathrm{pN}$ in Fig.~\ref{g0g0.02xu}(b) and (h) due to the decrease of the total effective potential function $V_{\text {eff }}(X)$ (the detailed calculation process is shown in the Appendix A) by the high-frequency signal $F$, the mechanical oscillations switch between the two potential wells. The system responses experience oscillations with larger amplitudes, resulting in amplified signals at the frequency $\omega$. When $G = 0.02 \omega_0$, the situation of $F = 316.3 \mathrm{pN}$ is similar to that when $F = 53.9 \mathrm{pN}$, in which case there are two potential wells but with a much lower effective potential compared to the original potential function ${V_0}(X)$, resulting in large-amplitude oscillations between the two wells [see Fig.~\ref{g0g0.02xu}(k)]. When the amplitude of the high-frequency signal further increases, that is, $F = 120 \mathrm{pN}$ and $335.8 \mathrm{pN}$, the mechanical mode primarily oscillates within the left potential well with a decreased amplification effect in Fig.~\ref{g0g0.02xu}(i) and (l). In conclusion, in the presence of DOPA, the high-frequency mechanical force enhances the system response to the weak signal by effectively lowering the potential barrier.

\begin{figure}[t]
\centerline{
\includegraphics[width=8.6cm, height=3.8cm, clip]{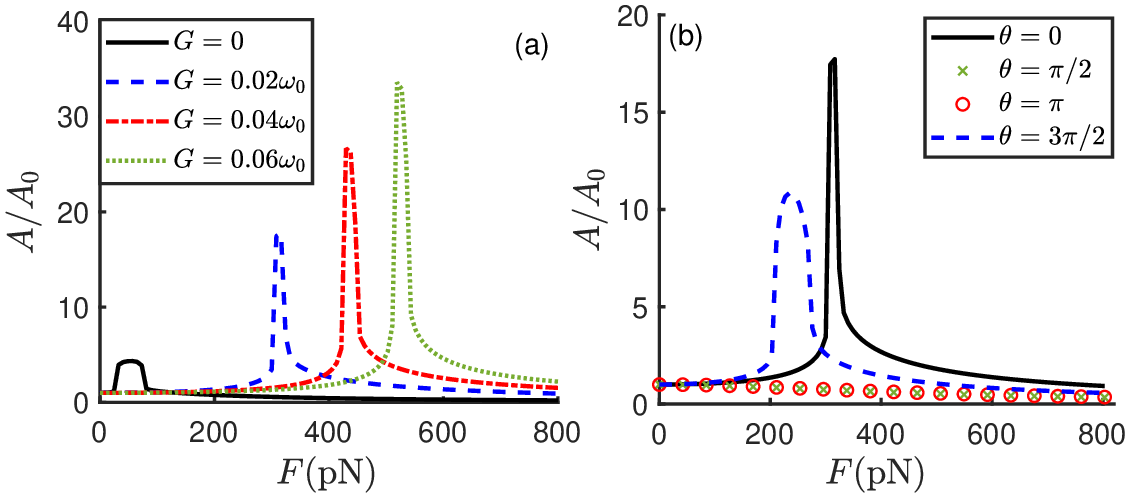}}
\caption{$A/A_{0}$ with different DOPA conditions ($\omega _g=2{\omega _m}$). The relative response amplitude $A/A_{0}$ of the system is given by Eq.~(\ref{q}) as a function of the amplitude of the high-frequency force $F$ for different (a) strength $G$ of the DOPA with $\theta = 0$, and (b) phase $\theta $ with $G = 0.02 \omega_0$. Other parameters are the same as  Fig.~\ref{sgzfig2}.}\label{sgzfig3}
\end{figure}

To explore the role of the DOPA in this resonator, we illustrate the relative response amplitude $A/A_{0}$ of the system versus the amplitude of the high-frequency force $F$ with different strength $G$ and phase $\theta $ of the DOPA when the frequency of high-frequency mechanical signal is set to $\Omega=510.5 \omega$ in Fig.~\ref{sgzfig3}. As shown in Fig.~\ref{sgzfig3}(a), we find when the strength $G$ of the DOPA increases from $0$ to $G = 0.06 \omega_0$, $A/A_{0}$ can be gradually enhanced. Significantly, when the strength $G = 0$, the amplification of the weak signal is significantly weaker than $G = 0.02 \omega_0$, $0.04 \omega_0$, and $0.06 \omega_0$. Figure~\ref{sgzfig3}(b) shows that the relative response amplitude $A/A_{0}$ of the system can also be tuned by manipulating $\theta $. Compared to the response at $\theta  = 0$, under the condition of maintaining a strength $G = 0.02 \omega_0$, $\theta = \pi /2$, $\theta  = \pi $, and $\theta  = 3 \pi /2$ result in lower relative response amplitude $A/A_{0}$. This result reveals the potential application of phase adjustment in optimizing amplification.

\begin{figure*}[t]
\centerline{
\includegraphics[width=17.5cm, height=5.5cm, clip]{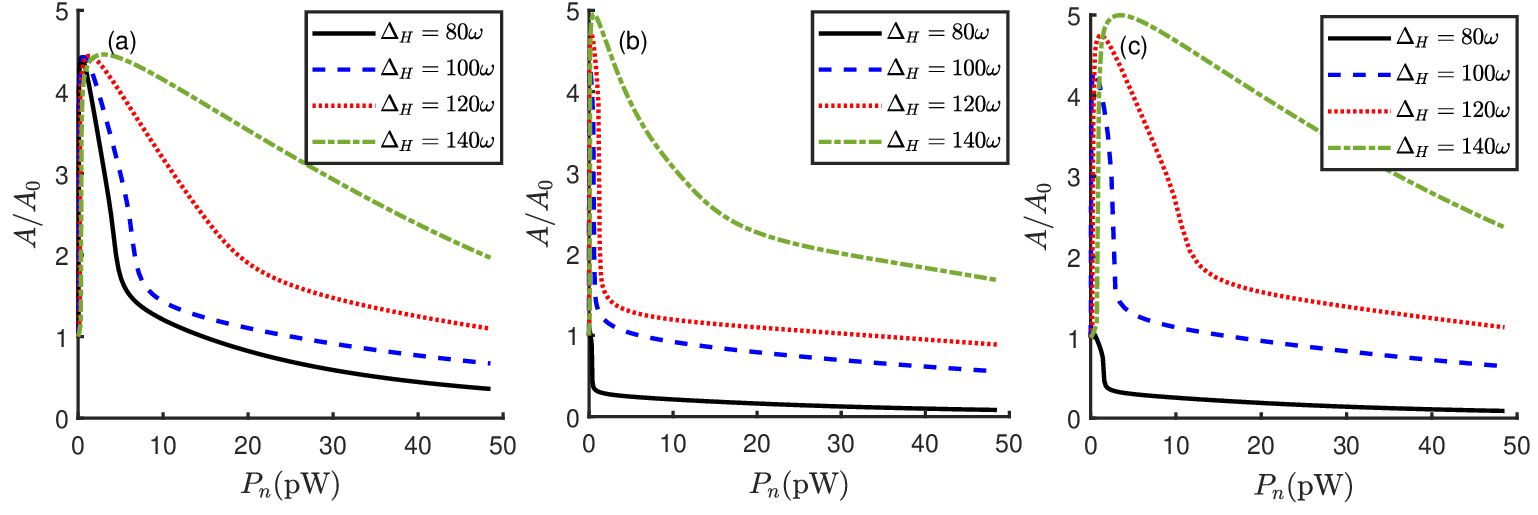}}
\caption{Impact of $\Delta_H$ on system amplification ($\omega_g=\omega_m + \omega_n$). The figure shows the relative response amplitude $A/A_{0}$ of the system determined by Eq.~(\ref{q}) as a function of the power $P_{n}$ of the optical signal for different frequency detunings $\Delta_H = 80 \omega, 100 \omega, 120 \omega$, and $140 \omega$, where the strength and phase of the probe field of the DOPA are fixed as (a) $G = 0$, $\theta  = 0$; (b) $G = 0.02 \omega_0$, $\theta = 0$; (c) $G = 0.02 \omega_0$, $\theta = 3\pi /2$. Other parameters are the same as  Fig.~\ref{sgzfig2}.}\label{sgzfig6}
\end{figure*}

\subsection{Influences of the DOPA with $\omega_g=\omega_m + \omega_n$ on the amplification of weak low-frequency mechanical signal by an optical signal}
With the capability of optomechanical systems interacting with both mechanical and optical signals, we would like to investigate the possibility of utilizing an optical signal as a high-frequency modulation signal to amplify the weak mechanical signal. Therefore, we replace the high-frequency mechanical force $F$ with an optical signal $\varepsilon _n$ and the frequency $\omega _n$ to drive the optical cavity. In this case, the total Hamiltonian~(\ref{H1}) of the system at the frequency ${{\omega _m}}$ is changed as
\begin{small}
\begin{equation}
\begin{aligned}
{{\hat H}} =& \frac{1}{2}m\omega _0^2{{\hat x}^2} + \frac{{{{\hat p}^2}}}{{2m}} +  \hbar\Delta_0 {{\hat a}^\dag }\hat a - \hbar \xi{{\hat a}^\dag }\hat a\hat x \\
&- i\hbar \sqrt \kappa  {\varepsilon _m}( {\hat a - {{\hat a}^\dag }} ) - i\hbar \sqrt \kappa  {\varepsilon _n}( {{e^{i\Delta_H t}}\hat a - {e^{ - i\Delta_H t}}{{\hat a}^\dag }} ) \\
&+ i\hbar G({{\hat a}^{\dag 2}}{e^{ - i\Delta_H t}}{e^{i\theta }} - H.c.) + {f}\cos ( {{\omega}t} )\hat x,\label{H2}
\end{aligned}
\end{equation}
\end{small}where $\Delta_H=\omega_{n}-\omega_{m}$ is the frequency detuning between the optical fields $\varepsilon _n$ and $\varepsilon _m$. We assume that this DOPA is excited by a pump field with the frequency $\omega _g={{\omega _m} + {\omega _n}}$ \cite{Liu99033822}, while $\omega _g=2{\omega _m}$ will be discussed in other sections.

To satisfy the condition of VR, which is the large frequency difference between two signals, we assume $\Delta_H \gg \omega$. The corresponding equations of motion for system variables are given by

\vspace{-1\baselineskip}
\begin{align}
\dot \beta  =&  - ( {i\Delta_0  + \frac{\kappa }{2}} )\beta  + i\xi\beta x + \sqrt \kappa  {\text{(}}{\varepsilon _m} + {\varepsilon _n}{e^{ - i\Delta_H t}}) \nonumber\\
&+ 2G{e^{i\theta }}{e^{ - i\Delta_H t}}{\beta ^*}
,\label{Heq7} \\
\dot x =& \frac{p}{m}
,\label{Heq8} \\
\dot p =&  - {\gamma}p - m\omega _0^2x +\hbar \xi|\beta {|^2} - {f}\cos ( {{\omega}t} )
.\label{Heq9}
\end{align}
Similarly, we can get the steady-state solution
\begin{equation}
\begin{aligned}
\beta  =& \frac{{8G\sqrt \kappa  {e^{i\theta  - i{\Delta _H}t}}( {{\varepsilon _m} + {\varepsilon _n}{e^{i{\Delta _H}t}}} )}}{b} \\
&- \frac{{2\sqrt \kappa  ( {2ia - \kappa } )( {{\varepsilon _m} + {\varepsilon _n}{e^{ - i{\Delta _H}t}}} )}}{b}
.\label{stableoptical}
\end{aligned}
\end{equation}
By substituting Eq.~(\ref{stableoptical}) into Eq.~(\ref{Heq9}), we arrive at the following equation of motion merely for the mechanical mode
\begin{widetext}
\begin{align}
\ddot x + {\gamma}\dot x =&  - \omega _0^2x - \frac{{{f}\cos \left( {{\omega}t} \right)}}{m}\nonumber\\
&+ \frac{{R[{e^{i\theta }}(O + Q{e^{2i{\Delta _H}t}}) + ({\varepsilon _m}{e^{i{\Delta _H}t}} + {\varepsilon _n})(\kappa  - 2ia)][O + Q + ({\varepsilon _m} + {\varepsilon _n}{e^{i{\Delta _H}t}})(\kappa  - 2ia)]}}{{{b^2}m}}
,\label{x2}
\end{align}
\end{widetext}
where $R = 4{e^{ - i(\theta  + 2{\Delta _H}t)}}\kappa \hbar \xi$, $O = 4{e^{i{\Delta _H}t}}{\varepsilon _m}G$, and $Q = 4{\varepsilon _n}G$. To see the influences of the DOPA on the enhancement of weak signal, the relative response amplitude $A/A_{0}$ of the system to the weak signal $f$ is investigated as a function of the power of the optical signal $P_{n}\left(P_{n}=\varepsilon _n^{2}  \hbar\omega_{n} / 2\right)$ at a few different frequencies ($\Delta_H=80 \omega, 100 \omega, 120 \omega$, and $140 \omega$) in Fig.~\ref{sgzfig6}. As shown in Fig.~\ref{sgzfig6}(a), in the absence of the DOPA, the relative response amplitude $A/A_{0}$ exhibits a maximum as the power of the optical signal $P_{n}$ varies, which is a typical feature of VR. In our model, the weak mechanical signal can be enhanced by an optical signal, whose control is usually more flexible than that of a high-frequency mechanical signal in experiments. We find in Fig.~\ref{sgzfig6}(b) that when $G = 0.02 \omega_0$, the relative response amplitude exhibits both increasing and decreasing trends. The peak of $A/A_{0}$ increases when $\Delta_H$ is larger (see red dotted line and green dash-dotted line in Fig.~\ref{sgzfig6}(b)). But the peak of $A/A_{0}$ decreases when $\Delta_H=80 \omega$ and $100 \omega$. In Fig.~\ref{sgzfig6}(c), when the phase $\theta $ of the DOPA increases from 0 to $3\pi /2$, $A/A_{0}$ still increases when $\Delta_H=120 \omega$ and $140 \omega$. In Fig.~\ref{sgzfig6}, we show that the amplification of the weak signal is sensitive to both the strength $G$ and phase $\theta $ changing of the DOPA. When $\Delta_H$ is larger, the influences of the $G$ and $\theta $ on the relative response amplitude $A/A_{0}$ become more obvious.

\begin{figure}[t]
\centerline{
\includegraphics[width=8.8cm, height=3.8cm, clip]{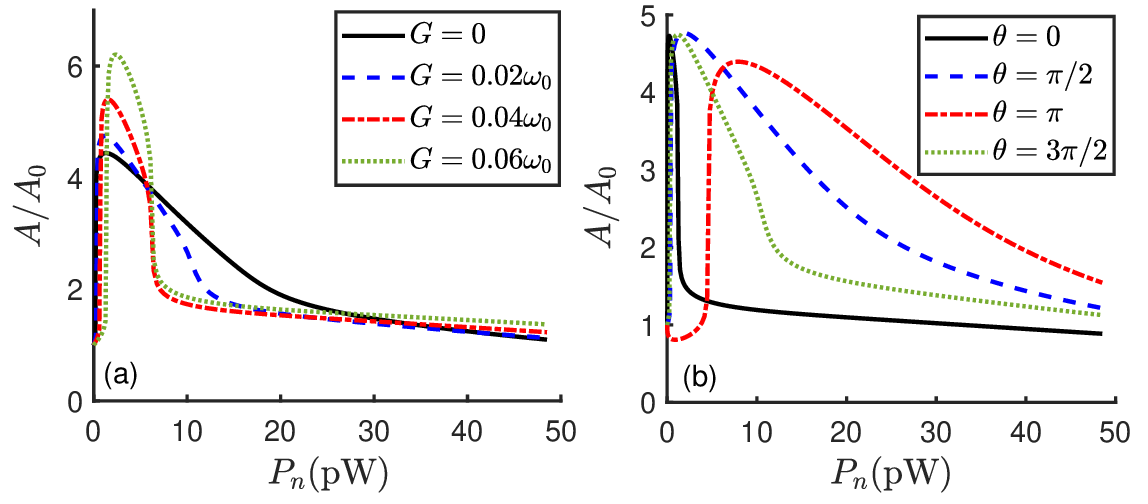}}
\caption{$A/A_{0}$ with different DOPA conditions ($\omega_g=\omega_m + \omega_n$). The relative response amplitude $A/A_{0}$ of the system is calculated by Eq.~(\ref{q}) as a function of the power $P_{n}$ of the optical signal for different (a) strength $G$ of the DOPA with $\theta = 3\pi /2$, and (b) phase $\theta $ with $G = 0.02 \omega_0$. Other parameters are the same as  Fig.~\ref{sgzfig2}.}\label{sgzfig7}
\end{figure}

\begin{figure}[t]
\centerline{
\includegraphics[width=8.9cm, height=6.6cm, clip]{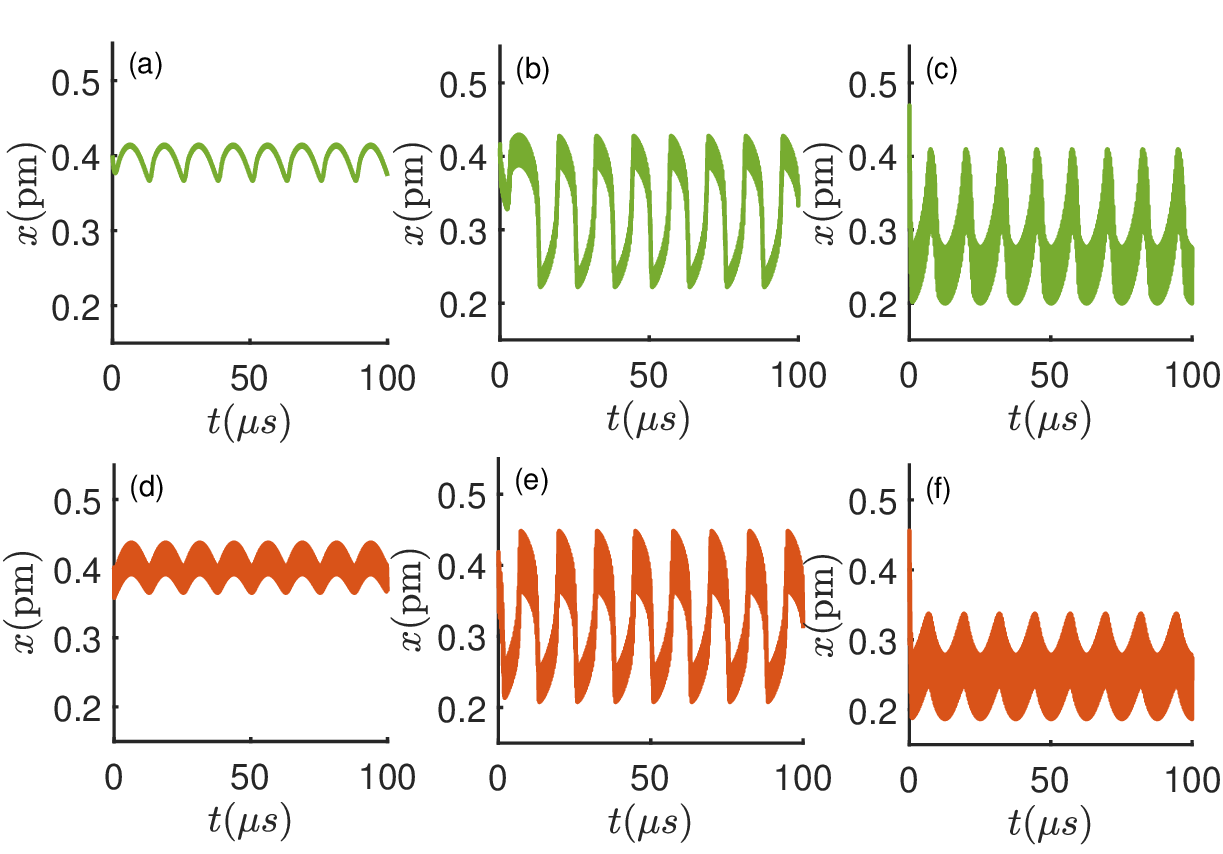}}
\caption{The mechanical position $x$ in Eqs.~(\ref{Heq7})-(\ref{Heq9}) as a function of the time $t$ for (a)-(c) $P_{n}=0.005 \mathrm{pW},1.11 \mathrm{pW}$, and $13.655 \mathrm{pW}$ with $G = 0$, (d)-(f) $P_{n}=0.005 \mathrm{pW},2.4 \mathrm{pW}$, and $13.655 \mathrm{pW}$ with $G = 0.06 \omega_0$, where $\Delta_H=120 \omega$ ($\omega_g=\omega_m + \omega_n$). Other parameters are the same as  Fig.~\ref{sgzfig2}.}\label{sgzfig8x}
\end{figure}

In Fig.~\ref{sgzfig7}, we discuss the influences of the DOPA on the relative response amplitude when the frequency of an extra optical signal takes $\Delta_H= 120 \omega$. We find when the strength $G$ of the DOPA increases from $0$ to $G = 0.06 \omega_0$, $A/A_{0}$ can be significantly enhanced in Fig.~\ref{sgzfig7}(a). The relative response amplitude can obtain obvious amplification when the power of the optical signal value is small. In Fig.~\ref{sgzfig7}(b), $A/A_{0}$ as a function of the power of the optical signal $P_{n}$ for the different $\theta $ at $G = 0.02 \omega_0$ is plotted. In detail, the relative response amplitude $A/A_{0}$ decreases when $\theta = \pi$. $A/A_{0}$ remains almost constant when $\theta$ takes other parameters.

To illustrate the dynamical behaviors of the system under different modulation strengths, in Fig.~\ref{sgzfig8x}(a)-(f) we present the time evolution of $x(t)$ at several representative points in the curves $G = 0$ and $G = 0.06 \omega_0$ of Fig.~\ref{sgzfig7}(a). For a very weak optical signal $\left(P_{n}=0.005 \mathrm{pW}\right)$, $x(t)$ experiences low-amplitude oscillations around one stable state in Fig.~\ref{sgzfig8x}(a) and (d). For a moderate optical signal, i.e., $P_{n}=1.11 \mathrm{pW}$ and $2.4 \mathrm{pW}$, $x(t)$ oscillates with a much larger amplitude between two stable states. Corresponding to the values of $A/A_{0}$, the oscillation amplitude with $G = 0.06 \omega_0$ is a little larger than $G = 0$. For the situation of the larger amplitude, i.e., $P_{n}=13.655 \mathrm{pW}$, $A/A_{0}$ for $G = 0$ is higher than that at $G = 0.06 \omega_0$. This exploration corresponds to the variation in the relative response amplitude $A/A_{0}$ in Fig.~\ref{sgzfig7}(a), proving that the DOPA has influences on the amplification.

\begin{figure*}[t]
\centerline{
\includegraphics[width=17.6cm, height=5.8cm, clip]{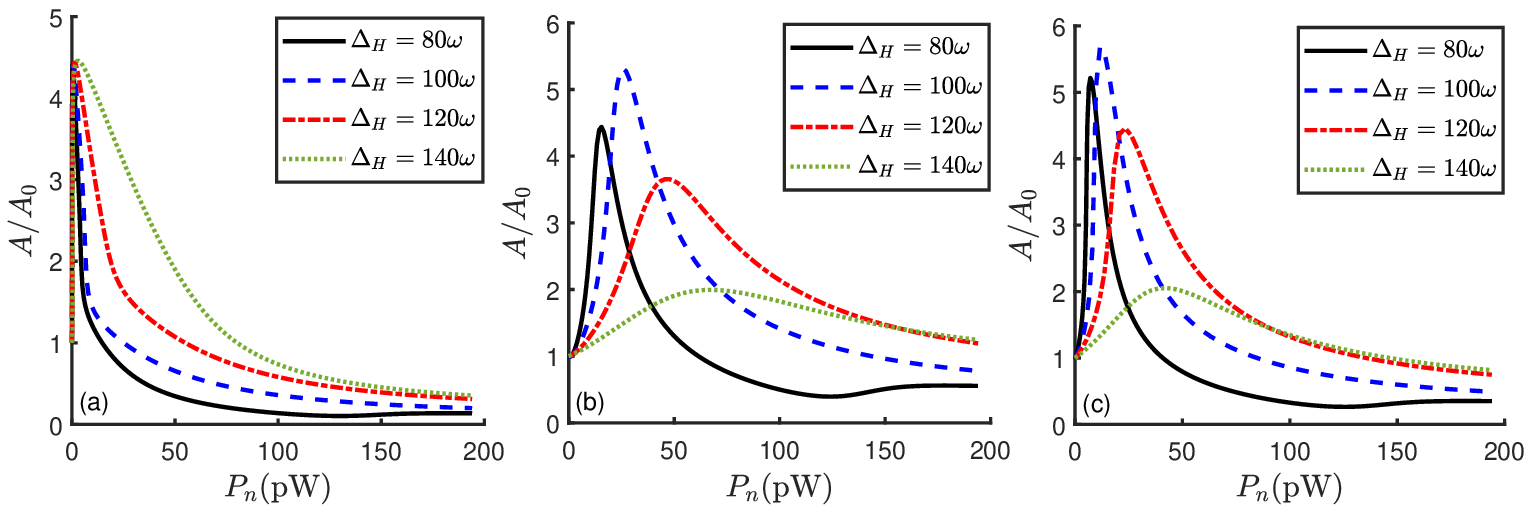}}
\caption{Impact of $\Delta_H$ on system amplification ($\omega _g=2{\omega _m}$). The figure shows the relative response amplitude $A/A_{0}$ of the system given by Eq.~(\ref{q}) as a function of the power $P_{n}$ of the optical signal for different values of $\Delta_H$, where the strength and phase of the probe field of the DOPA take (a) $G = 0,\theta  = 0$; (b) $G = 0.02 \omega_0$, $\theta = 0$; (c) $G = 0.02 \omega_0$, $\theta = 3\pi /2$.  Other parameters are the same as  Fig.~\ref{sgzfig2}.}\label{sgzfig11}
\end{figure*}

\subsection{Influences of changing the driving frequency of the DOPA with $\omega_g=2{\omega _m}$ on the amplification}
The optical degenerate parametric amplifier (a second-order optical crystal in nature) can generate pairs of down-converted photons and show nearly perfect single or dual squeezing \cite{Nation841,Gerry,Li100023838,Clerk821155,Leghtas347853,Shen100023814}. As is well-known, placing a DOPA pumped by an external laser in the optomechanical cavity can modulate the optomechanical coupling, which leads to optical amplification directly  \cite{Adamyan92053818}. We can discuss the influences of different pump frequencies of the DOPA on the system and compare the amplification of the weak signal $f$ in both cases. Now, we vary the frequency of the DOPA so that the DOPA is excited by a pump field with the frequency $\omega _g=2{\omega _m}$ \cite{Li100023838}. The pump photon with frequency $\omega_g=2{\omega _m}$ is down-converted into an identical pair of photons with frequency ${\omega _m}$ after passing through the second-order nonlinearity crystal. In this case, the total Hamiltonian of the system in the rotating frame at the frequency ${{\omega _m}}$ is given by
\begin{small}
\begin{equation}
\begin{aligned}
{{\hat H}} =& \frac{1}{2}m\omega _0^2{{\hat x}^2} + \frac{{{{\hat p}^2}}}{{2m}} +  \hbar\Delta_0 {{\hat a}^\dag }\hat a -  \hbar\xi {{\hat a}^\dag }\hat a\hat x \\
&- i \hbar\sqrt \kappa  {\varepsilon _m}( {\hat a - {{\hat a}^\dag }} ) - i\hbar \sqrt \kappa  {\varepsilon _n}( {{e^{i\Delta_H t}}\hat a - {e^{ - i\Delta_H t}}{{\hat a}^\dag }} )\\
&+ i\hbar G({{\hat a}^{\dag 2}}{e^{i\theta }} - H.c.) + {f}\cos ( {{\omega}t} )\hat x
,\label{H3}
\end{aligned}
\end{equation}
\end{small}
which leads to the equations of motion

\vspace{-1\baselineskip}
\begin{align}
\dot \beta  =&  - ( {i\Delta_0  + \frac{\kappa }{2}} )\beta  + i\xi \beta x + \sqrt \kappa  {\text{(}}{\varepsilon _m} + {\varepsilon _n}{e^{ - i\Delta_H t}}) \nonumber\\
&+ 2G{e^{i\theta }}{\beta ^*}
,\label{Heq13}\\
\dot x =& \frac{p}{m}
,\label{Heq14}\\
\dot p =&  - {\gamma}p - m\omega _0^2x + \hbar \xi |\beta {|^2} - {f}\cos ( {{\omega}t} )
,\label{Heq15}
\end{align}
and corresponding the steady-state solution of the optical mode
\begin{equation}
\begin{aligned}
\beta  =& \frac{{8G\sqrt \kappa  {e^{i\theta }}( {{\varepsilon _m} + {\varepsilon _n}{e^{i{\Delta _H}t}}} )}}{b} \\
&- \frac{{2\sqrt \kappa  ( {2ia - \kappa } )( {{\varepsilon _m} + {\varepsilon _n}{e^{ - i{\Delta _H}t}}} )}}{b}
.\label{betacwentaijie1}
\end{aligned}
\end{equation}

To illustrate the different influences on the DOPA excited by a pump field with frequency $\omega_g=2{\omega _m}$, the relative response amplitude $A/A_{0}$ of the system is investigated as a function of the power $P_{n}$ of the optical signal in Fig.~\ref{sgzfig11}. In Fig.~\ref{sgzfig11}(a), in the absence of the DOPA, we find that resonance peaks are present for all four curves, signaling the occurrence of the VR phenomenon. Obviously, the results show that the weak force can be amplified by adding an optical signal when the pump field has a frequency $\omega_g=2{\omega _m}$. Significantly, as the frequency of the extra optical signal increases, the peak value of the relative response amplitude $A/A_{0}$ remains almost constant. In Fig.~\ref{sgzfig11}(b), the relative response amplitude exhibits a trend of both increasing and decreasing variations. The relative response amplitude $A/A_{0}$ gets larger in the presence of the DOPA for $\Delta_H= 80 \omega$ and $100 \omega$. To be specific, for $G = 0.02 \omega_0$, $\theta  = 0$, and $\Delta_H= 80 \omega$, $A/A_{0}$ can increase from 4.41 to 4.43. When the optical signal frequency increases to $\Delta_H= 100 \omega$, $A/A_{0}$ increases from 4.42 to 5.31. However, as the frequency $\Delta_H$ continues to increase to higher values such as $\Delta_H = 120 \omega$ and $140 \omega$, the relative response amplitude $A/A_{0}$ decreases. While it is interesting that we can see with tuning $\theta $ changing from $\theta = 0$ to $\theta  = 3\pi /2$, the peaks of all four curves increase in Fig.~\ref{sgzfig11}(c). We find that the relative response amplitude $A/A_{0}$ exhibits strong sensitivity to variations in both the strength $G$ and the phase $\theta $ of the DOPA.

\begin{figure}[t]
\centerline{
\includegraphics[width=8.4cm, height=4cm, clip]{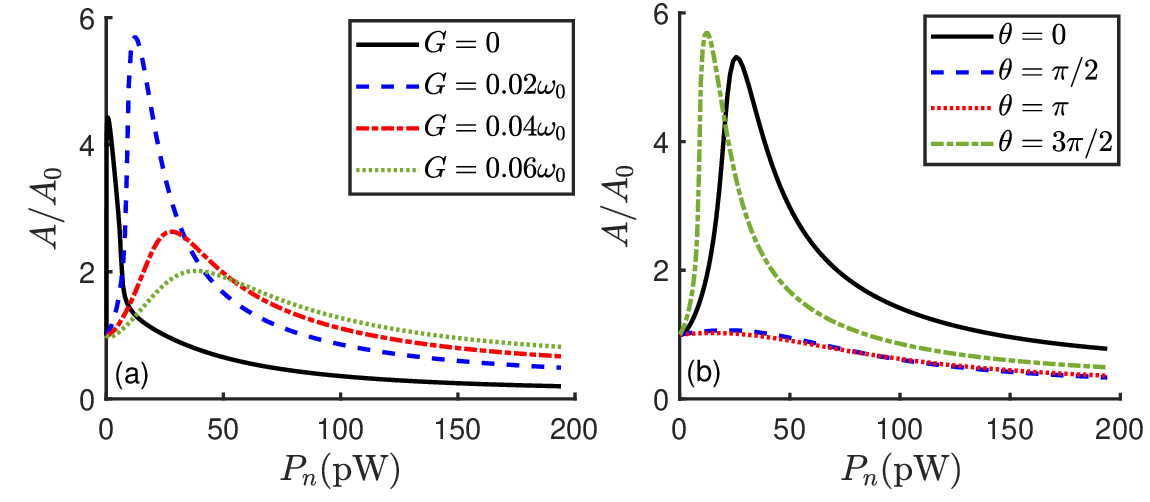}}
\caption{$A/A_{0}$ with different DOPA conditions ($\omega _g=2{\omega _m}$). The relative response amplitude $A/A_{0}$ of the system is solved by Eq.~(\ref{q}) as a function of the power $P_{n}$ of the optical signal for different (a) strength $G$ of the DOPA with $\theta = 3\pi /2$, and (b) phase $\theta $ with $G = 0.02 \omega_0$. Other parameters are the same as  Fig.~\ref{sgzfig2}.}\label{sgzfig12}
\end{figure}

Next, we discuss the influences of the DOPA on the relative response amplitude $A/A_{0}$ of the system when the frequency of an extra optical signal takes $\Delta_H= 100 \omega$. In Fig.~\ref{sgzfig12}, we show that both $G$ and $\theta $ change the peak of $A/A_{0}$. In Fig.~\ref{sgzfig12}(a), as $G$ increases, the position of the peak shifts to the right, i.e., a larger value of $P_{n}$ is needed to bring $A/A_{0}$ to its maximum. In particular, when $G = 0.02 \omega_0$, $A/A_{0}$ appears as a local maximum at $P_{n} = 5.65 \mathrm{pW}$. In the case of $G = 0.02 \omega_0$ in Fig.~\ref{sgzfig12}(b), we find that compared to $\theta  = 0$, both $\theta = \pi /2$ and $\theta  = \pi $ result in a lower relative response amplitude $A/A_{0}$ of the system, but $A/A_{0}$ is amplified when $\theta  = 3 \pi /2$, which obtains the maximum value of 5.7.

In summary, when the frequency of the DOPA takes $\omega_g=2{\omega _m}$, the DOPA is significant for the amplification of the lower frequency of the optical signal. Compared to the case where $\omega_g={{\omega _m} + {\omega _n}}$, the DOPA can enhance the peak of the higher frequency of the optical signal.

\section{Amplification of a weak low-frequency mechanical signal in non-Markovian systems}
\subsection{Amplification of a weak low-frequency mechanical signal by the DOPA in non-Markovian systems utilizing a high-frequency mechanical signal as modulation signal}
When the system interacts with the environment, the environmental influences on the system dynamics show the decay or the backflow oscillation of photons coming from the environment. To be specific, the decay behavior corresponds to the Markovian approximation, whereas the backflow oscillation demonstrates non-Markovian effects of the environment acting on the system \cite{breuer2002,breuer1032104012009,Vega015001}. In the previous section, we have studied weak-force amplification in cavity-optomechanical systems under the Markovian approximation. In this section, we investigate the influences of non-Markovian effects on the amplification of a weak low-frequency mechanical signal by a high-frequency mechanical signal of the DOPA. For this purpose, we consider that the cavity couples with the non-Markovian environment, which consists of a series of boson modes (eigenfrequency ${\omega _k}$) \cite{Xiong2019100,Xu201082,hoeppe1080436032012,madsen1062336012011,Guo2021126,Li2022129,Uriri2020101,Liu2020102,Anderson199347,Vega015001}. After making a unitary transformation via $\hat U\hat H_1{{\hat U}^\dag } - i \hbar\hat U \partial {\hat U^\dag }/\partial t $ with ${{\hat U}}(t) = \exp [ - i{\omega _m}t({{\hat a}^\dag }\hat a + \sum\nolimits_k {\hat b_k^\dag } {{\hat b}_k})]$ to the rotating frame and assuming $\omega_g=2{\omega _m}$, the total Hamiltonian (\ref{H1}) is changed to

\vspace{-1\baselineskip}
\begin{align}
{{\hat H}} =& \ \frac{1}{2}m\omega _0^2{{\hat x}^2} + \frac{{{{\hat p}^2}}}{{2m}} +  \hbar\Delta_0 {{\hat a}^\dag }\hat a - \hbar \xi {{\hat a}^\dag }\hat a\hat x \nonumber\\
&- i \hbar\sqrt \kappa  {\varepsilon _m}( {\hat a - {{\hat a}^\dag }} ) + i\hbar G({{\hat a}^{\dag 2} }{e^{i\theta }} - H.c.) \nonumber\\
&+ {f}\cos ( {{\omega}t} )\hat x + {F}\cos ( {{\Omega}t} )\hat x \nonumber\\
&+ \hbar \sum\limits_k {{\Delta _k}\hat b_k^\dag {{\hat b}_k}}  + i\hbar \sum\limits_k {({g_k}\hat a} \hat b_k^\dag  - g_k^*{{\hat a}^\dag }{{\hat b}_k}),\label{H4}
\end{align}
where ${\Delta _k} = {\omega _k} - {\omega _m}$ defines the detuning of $k$th mode (eigenfrequency ${\omega _k}$) of the environment from the driving field. ${{\hat b}_k}(\hat b_k^\dag )$ is the annihilation (creation) operator of the environment. ${{g_k}}$ denotes the coupling coefficient between the cavity and the environment. The Heisenberg operators in Eq.~(\ref{H4}) satisfy

\vspace{-1\baselineskip}
\begin{align}
\frac{d}{{dt}}\hat a(t) =&  - i\Delta_0 \hat a(t) + i\xi \hat a(t)\hat x(t) + 2G{e^{i\theta }}{{\hat a}^\dag }(t)\nonumber\\
&+ \sqrt \kappa  \left( {{\varepsilon _m} + {{\hat a}_{{\rm{in}}}}} \right) - \sum\limits_k {g_k^*{{\hat b}_k}(t)}
,\label{Heq19}\\
\frac{d}{{dt}}\hat x(t) =& \frac{{\hat p(t)}}{m}
,\label{Heq20}\\
\frac{d}{{dt}}\hat p(t) =&  - {\gamma}\hat p(t) - m\omega _0^2\hat x(t) + \hbar \xi {{\hat a}^\dag }(t)\hat a(t) \nonumber\\
&- {f}\cos \left( {{\omega}t} \right) - {F}\cos \left( {{\Omega}t} \right)
,\label{Heq21}\\
\frac{d}{{dt}}{{\hat b}_k}(t) =&  - i{\Delta _k}{{\hat b}_k}(t) + {g_k}\hat a(t)
.\label{Heq22}
\end{align}
Under the condition of strong driving and weak coupling between the optical mode and mechanical mode, we approximately ignore the quantum fluctuations. With mean-field amplitudes $\hat{a} \rightarrow \beta$, $\hat{x} \rightarrow x$, $\hat{p} \rightarrow p$, and ${{\hat b}_k} \rightarrow {b_k}$ \cite{Vitali98030405,Aspelmeyer861391}, the equations of motion for the mean-field amplitudes of the system operators give

\vspace{-1\baselineskip}
\begin{small}
\begin{align}
\dot \beta  =&  - i\Delta_0 \beta  + i\xi \beta x + \sqrt \kappa  {\varepsilon _m} + 2G{e^{i\theta }}{\beta ^*} - \sum\limits_k {g_k^*{b_k}}
,\label{Heq23}\\
\dot x =& \frac{p}{m}
,\label{Heq24}\\
\dot p =&  - {\gamma}p - m\omega _0^2x + \hbar \xi |\beta {|^2} - {f}\cos \left( {{\omega}t} \right) - {F}\cos \left( {{\Omega}t} \right)
,\label{Heq25}\\
{{\dot b}_k} =&  - i{\Delta _k}{b_k} + {g_k}\beta
.\label{Heq26}
\end{align}
\end{small}
By solving Eq.~(\ref{Heq26}), we obtain
\begin{equation}
\begin{aligned}
{b_k}(t) = {b_k}(0){e^{ - i{\Delta _k}t}} + {g_k}\int_0^t {d\tau } \beta (\tau ){e^{ - i{\Delta _k}(t - \tau )}}
.\label{bk1}
\end{aligned}
\end{equation}
The first term on the right-hand side of Eq.~(\ref{bk1}) represents the freely propagating parts of the environmental field, while the second term describes the influences of the environment on the cavity. Substituting Eq.~(\ref{bk1}) into Eq.~(\ref{Heq23}), we get an integro-differential equation
\begin{equation}
\begin{aligned}
\dot \beta  =&  - i\Delta_0 \beta  + i\xi \beta x + \sqrt \kappa  {\varepsilon _m} + 2G{e^{i\theta }}{\beta ^*} \\
&+ {{Z}}(t) - \int_0^t {d\tau } \beta (\tau )C(t - \tau )
,\label{jifenweifenfangchengsuanfu}
\end{aligned}
\end{equation}where ${{\hat Z} }(t) =  - \sum\nolimits_k {g_k^*{{\hat b}_k}(0){e^{ - i{\Delta _k}t}}}  =   \int_{ - \infty }^\infty  {d\tau {h ^*}} (t - \tau ){\hat a_{in}}(\tau )$. The cavity couples to the incoming and outgoing modes of the environment at both ends. We define here the input-field operator ${\hat a_{in}}(t) = \frac{1}{{\sqrt {2\pi } }}\sum\nolimits_k {{e^{ - i{\Delta _k}t}}{{\hat b}_k}(0)} $ and impulse response function $h (t) = \frac{-1}{{\sqrt {2\pi } }}\sum\nolimits_k {{e^{i{\Delta _k}t}}{g_k}} $, or in the continuum $h (t - \tau ) = \frac{-1}{{\sqrt {2\pi } }}\int {{e^{i(\omega ' - \omega_m) (t - \tau )}}g(\omega ')} d\omega ' $, where we have made the replacement ${g_k} \to g(\omega ' )$.
${\hat a_{in}}(t)$ is the input field with zero means ${a_{in}}(t) = \langle {{{\hat a}_{in}}(t)} \rangle  = 0$ for the environment initialization in the vacuum state, which leads to ${{ Z} }(t) =\langle {\hat Z(t)} \rangle = 0$. The correlation function in Eq.~(\ref{jifenweifenfangchengsuanfu}) is given by
\begin{equation}
\begin{aligned}
C(t) = \sum_k {{{| {{g_k}} |}^2}{e^{ - i{\Delta _k}t}}}  = \int {J(\omega ' ){e^{ - i(\omega ' - \omega_m) t}}d\omega ' },
\label{ft}
\end{aligned}
\end{equation}where $J(\omega ' ) = \sum\nolimits_k {|{g_k}{|^2}} \delta (\omega '  - {\omega _k})$ represents the spectral density of the environment. $C(t)$ in Eq.~(\ref{ft}) denotes the correlation function of the non-Markovian environment, whose detail derivation can be found in Appendix~\ref{APP.2}. The Lorentzian spectral density of the environment takes \cite{breuer2002,Shen1070537052023,shen1050237072022,Zhang1090337012024,Yang1090537122024,shen880338352013,zhang870321172013,diosi850341012012,xiong860321072012,shen950121562017,Jack630438032001}

\vspace{-1\baselineskip}
\begin{equation}
\begin{aligned}
J(\omega ' ) = \frac{{{\kappa}}}{{2\pi }}\frac{{\lambda^2}}{{\lambda^2 + {(\omega ' - \omega_m) ^2}}},
\label{spectral_density}
\end{aligned}
\end{equation}
which is implemented via all-optical setups \cite{Xiong2019100,Fanchini112210402} and pseudomode methods \cite{Barnett1997,Garraway552290,Tamascelli120030402}. With Eq.~(\ref{spectral_density}), we get $C(t - \tau ) = \frac{1}{2}{\kappa}{\lambda}{e^{ - {\lambda}\left| {t - \tau } \right|}}$ (representing a Gaussian Ornstein-Uhlenbeck process \cite{uhlenbeck368231930,gillespie5420841996,jing1052404032010}), which can also be realized through the pseudomode theory (see Appendix~\ref{APP.3}). The steady-state solution with the non-Markovian environment can be obtained from Eq.~(\ref{jifenweifenfangchengsuanfu}) as

\vspace{-1\baselineskip}
\begin{equation}
\begin{aligned}
\beta  = \frac{{2\sqrt \kappa  {\varepsilon _m}(4{e^{i\theta }}G - 2ia + \kappa )}}{b}
,\label{wentaijie4}
\end{aligned}
\end{equation}
where we can adiabatically eliminate the optical mode by substituting Eq.~(\ref{alphastable}) into the equations of motion for the mechanical mode in Eq.~(\ref{Heq6}), and then write the equation of motion merely for the mechanical mode as

\begin{small}
\begin{equation}
\begin{aligned}
\ddot x + \gamma \dot x =&  - \omega _0^2x - \frac{{f\cos ( {\omega t} )}}{m} - \frac{{F\cos ( {\Omega t} )}}{m}\\
&+ \frac{{c( { - 2ia + \kappa  + 4G{e^{i\theta }}} )( {2ia + \kappa  + 4G{e^{ - i\theta }}} )}}{{{b^2}m}}
.\label{x1}
\end{aligned}
\end{equation}
\end{small}

\begin{figure}[t]
\centerline{
\includegraphics[width=9cm, height=4cm, clip]{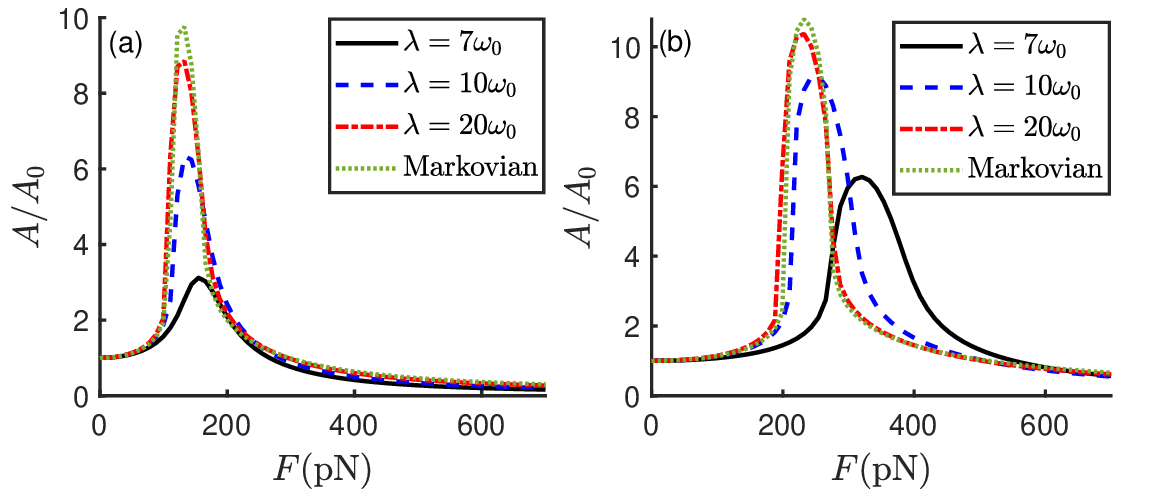}}
\caption{Effect of environmental spectrum width $\lambda$ ($\omega _g=2{\omega _m}$). The figure shows the relative response amplitude $A/A_{0}$ of the system solved by Eq.~(\ref{q}) as a function of $F$, which corresponds to the Markovian and non-Markovian environments with the different environmental spectrum width $\lambda$ and DOPA involvement ($G = 0.02 \omega_0$ and $\theta = 3\pi /2$). The high-frequency force frequency corresponds to (a) $\Omega=400 \omega$; (b) $\Omega=510.5 \omega$. Other parameters are the same as  Fig.~\ref{sgzfig2}.}\label{fmsgzfig16}
\end{figure}

In Fig.~\ref{fmsgzfig16}(a) with $\Omega=400 \omega$ and the presence of the DOPA for $G = 0.02 \omega_0$ and $\theta  = 3\pi /2$, we show the relative response amplitude $A/A_{0}$ of the system versus the amplitude of the high-frequency force $F$ with the different spectral width $\lambda$ of the environment. For a given spectral width of environment, decreasing from ${\lambda} = 20{\omega _0}$ to $7{\omega _0}$, we find from the figure that the relative response amplitude $A/A_{0}$ gradually decreases in the non-Markovian environment. From Fig.~\ref{fmsgzfig16}(b), when the frequency $\Omega=510.5 \omega$, the relative response amplitude $A/A_{0}$ also decreases in the non-Markovian environment. Compared to the Markovian environment, the relative response amplitude $A/A_{0}$ gradually decreases when the weak mechanical signal is tuned by a high-frequency mechanical signal.

\begin{figure}[t]
\centerline{
\includegraphics[width=9cm, height=4cm, clip]{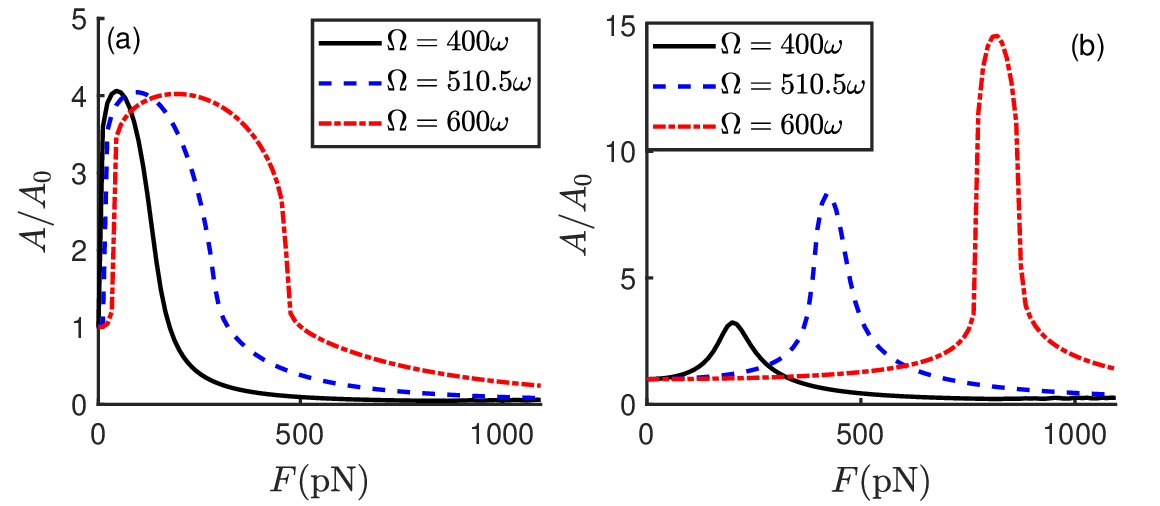}}
\caption{Impact of the modulation frequency $\Omega$ on amplification in the non-Markovian environment ($\omega _g=2{\omega _m}$). The relative response amplitude $A/A_{0}$ of the system is determined by Eq.~(\ref{q}) as a function of $F$, where the strength and phase of the probe field of the DOPA are fixed as (a) $G = 0,\theta = 0$; (b) $G = 0.02 \omega_0$, $\theta = 0$. Other parameters are the same as  Fig.~\ref{sgzfig2}.}\label{nonomega}
\end{figure}

\begin{figure*}[t]
\centerline{
\includegraphics[width=17.5cm, height=5.8cm, clip]{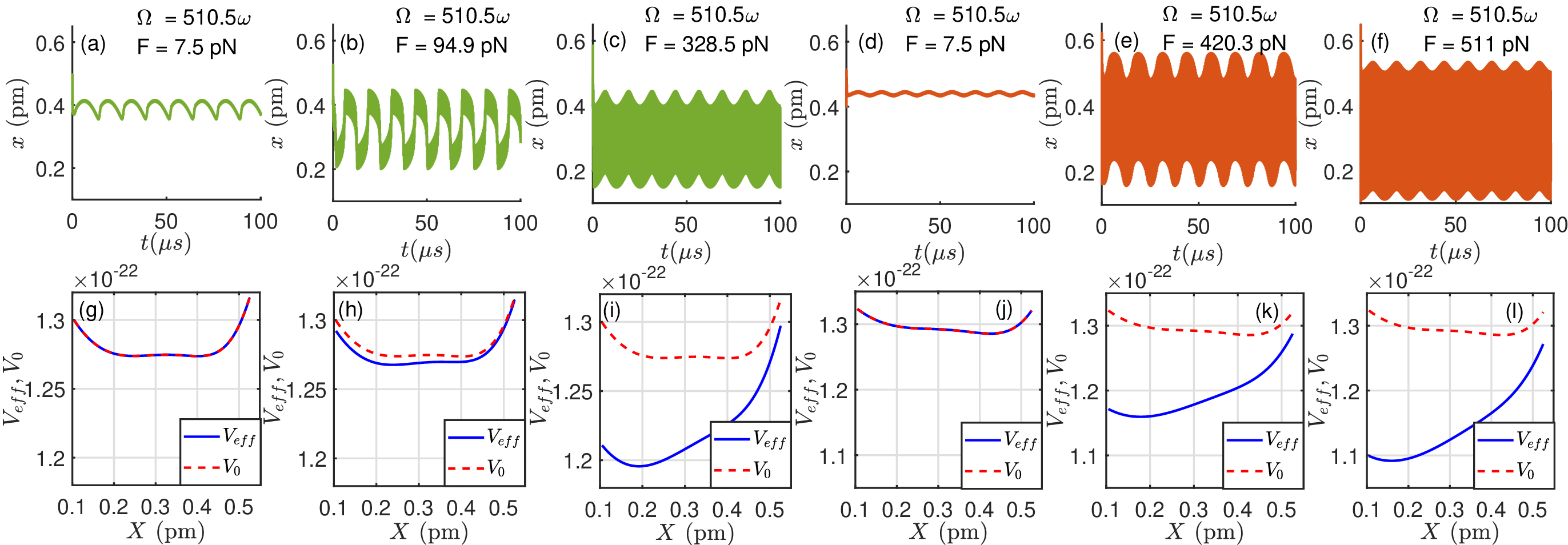}}
\caption{The phenomenon of enhancing a weak low-frequency mechanical signal ($\omega _g=2{\omega _m}$). The figure shows the mechanical position $x$ in Eqs.~(\ref{Heq23})-(\ref{Heq26}) as a function of the time $t$ at different high-frequency force $F$ for $\Omega=510.5 \omega$, where the strength and phase of the probe field of the DOPA take (a)-(c) $G = 0$, $\theta = 0$; (d)-(f) $G = 0.02 \omega_0$, $\theta = 0$; (g)-(l) The effective potential function $V_{\text {eff }}(X)$ corresponding to Fig.~\ref{nonxtandu}(a)-(f) is obtained according to Eq.~(\ref{ueff1}). $V_{0}(X)$ in Eq.~(\ref{u0}) is the potential function in the absence of $F$. Other parameters are the same as  Fig.~\ref{sgzfig2}.}\label{nonxtandu}
\end{figure*}

Similar to the Markovian environment, we also use three high-frequency forces $F$ with $\Omega = 400 \omega, 510.5 \omega$, and $600 \omega$ to study the influences of the DOPA on the enhancement of weak signal in the non-Markovian environment shown in Fig.~\ref{nonomega}. In Fig.~\ref{nonomega}(a), we discuss that the relative response amplitude $A/A_{0}$ varies with the high-frequency force $F$ without the participation of the DOPA (the strength $G = 0$ and phase $\theta  = 0$). Under the resonator stationary, all three curves exhibit resonance peaks, indicating the occurrence of the VR phenomenon. The resonance peak shifts towards larger values of $F$ as the modulation frequency $\Omega$ increases. As shown in Fig.~\ref{nonomega}(b), we present the changes in the relative response amplitude $A/A_{0}$ in the presence of the DOPA. Specifically, when $G = 0.02 \omega_0$, $\theta  = 0$, and $\Omega = 510.5 \omega$, $A/A_{0}$ can increase from 4 to 8.3. When the frequency of high-frequency mechanical signal $\Omega = 600 \omega$, $A/A_{0}$ also can increase from 4 to 14.6. But at the higher frequency of $\Omega = 400 \omega$, the relative response amplitude $A/A_{0}$ exhibits a decreasing trend. Different from the Markovian environment shown in Fig.~\ref{sgzfig2}(b), the weak signal can be enhanced under the high-frequency mechanical force $F$ under the action of the DOPA in the non-Markovian environment.

To deeply explore the physical reason behind $A/A_{0}-F$ curves in the presence of the DOPA, we respectively show the dynamics $x(t)$ at three points on $A/A_{0}-F$ curve of $\Omega = 510.5 \omega$, $G = 0$, $\theta = 0$ in Fig.~\ref{nonxtandu}(a)-(c) and $G = 0.02 \omega_0$, and $\theta = 0$ in Fig.~\ref{nonxtandu}(d)-(f). Moreover, we plot the corresponding effective potential function $V_{\text {eff }}(X)$ (the detailed calculation process is shown in the Appendix A) in Fig.~\ref{nonxtandu}(g)-(l). When the amplitude of the high-frequency force is low (e.g., $F=7.5 \mathrm{pN}$), the position of the mechanical system oscillates with a small amplitude around a stable state, which corresponds to the right well of the potential function in Fig.~\ref{nonxtandu}(g) and (j). When the amplitude rises to $F=94.9 \mathrm{pN}$ in Fig.~\ref{nonxtandu}(b) and (h), due to the decrease of the total effective potential function $V_{\text {eff }}(X)$ by the high-frequency signal $F$, this makes the mechanical oscillations switch between the two potential wells. The system responses experience oscillations with significantly larger amplitudes, resulting in significantly amplified signals at the frequency $\omega$. When $G = 0.02 \omega_0$, the situation of $F = 420.3 \mathrm{pN}$ is similar to that if $F = 94.9 \mathrm{pN}$, in which case there are two potential wells but with much lower effective potential compared to the original potential function ${V_0}(X)$, resulting in large-amplitude oscillation between two wells [see Fig.~\ref{nonxtandu}(k)]. When the amplitude of the high-frequency signal further increases, that is, $F = 328.5 \mathrm{pN}$ and $511 \mathrm{pN}$, the mechanical mode primarily oscillates within the left potential well with a decreased amplification effect and the low-frequency motion is nearly buried in the strong rapid oscillations in Fig.~\ref{nonxtandu}(i) and (l). In conclusion, in the presence of DOPA within the non-Markovian environment, the high-frequency mechanical force enhances the response of the system to the weak signal by effectively lowering the potential barrier.
\subsection{Amplification of weak low-frequency mechanical signal by the DOPA with $\omega _g={{\omega _m} + {\omega _n}}$ in non-Markovian systems utilizing an optical signal as modulation signal}
As optomechanical systems are capable of interacting with both mechanical and optical signals, we investigate the influences of non-Markovian effects on the amplification of a weak low-frequency mechanical signal by an optical signal of the DOPA. In this case, we assume that this DOPA is excited by a pump field with the frequency  $\omega _g={{\omega _m} + {\omega _n}}$ \cite{Liu99033822}. The total Hamiltonian of the system in the rotating frame at the laser frequency ${{\omega _m}}$ reads
\begin{small}
\begin{equation}
\begin{aligned}
{{\hat H}} =& \frac{1}{2}m\omega _0^2{{\hat x}^2} + \frac{{{{\hat p}^2}}}{{2m}} +  \hbar\Delta_0 {{\hat a}^\dag }\hat a - \hbar \xi {{\hat a}^\dag }\hat a\hat x \\
&- i \hbar\sqrt \kappa  {\varepsilon _m}( {\hat a - {{\hat a}^\dag }} ) - i \hbar\sqrt \kappa  {\varepsilon _n}( {{e^{i\Delta_H t}}\hat a - {e^{ - i\Delta_H t}}{{\hat a}^\dag }} ) \\
&+ i\hbar G({{\hat a}^{\dag 2} }{e^{ - i\Delta_H t}}{e^{i\theta }} - H.c.) + {f}\cos ( {{\omega}t} )\hat x \\
&+ \hbar \sum\limits_k {{\Delta _k}\hat b_k^\dag {{\hat b}_k}}  + i\hbar \sum\limits_k {({g_k}\hat a} \hat b_k^\dag  - g_k^*{{\hat a}^\dag }{{\hat b}_k})
,\label{H5}
\end{aligned}
\end{equation}
\end{small}
which causes

\vspace{-1\baselineskip}
\begin{align}
\dot \beta  =&  - i\Delta_0 \beta  + i\xi \beta x + \sqrt \kappa  {\rm{(}}{\varepsilon _m} + {\varepsilon _n}{e^{ - i\Delta_H t}})\nonumber\\
&+ 2G{e^{i\theta }}{e^{ - i\Delta_H t}}{\beta ^*}- \sum\limits_k {g_k^*{b_k}}
,\label{Heq31}\\
\dot x =& \frac{p}{m}
,\label{Heq32}\\
\dot p =&  - {\gamma}p - m\omega _0^2x + \hbar \xi |\beta {|^2} - {f}\cos ( {{\omega}t} )
,\label{Heq33}\\
{{\dot b}_k} =&  - i{\Delta _k}{b_k} + {g_k}\beta
,\label{Heq34}
\end{align}
where we have written the operators for their expectation values by the mean-field approximation. Solving Eq.~(\ref{Heq34}) leads to Eq.~(\ref{Heq31}) being
\begin{equation}
\begin{aligned}
\dot \beta  =& - i\Delta_0 \beta  + i\xi \beta x + \sqrt \kappa  {\rm{(}}{\varepsilon _m} + {\varepsilon _n}{e^{ - i\Delta_H t}})\\
&+ 2G{e^{i\theta }}{e^{ - i\Delta_H t}}{\beta ^*} - \int_0^t {d\tau } \beta (\tau )C(t - \tau )
.\label{jifenweifenfangcheng2}
\end{aligned}
\end{equation}

The corresponding steady-state solution containing non-Markovian effects can be written as
\begin{small}
\begin{equation}
\begin{aligned}
\beta  =& \frac{{2\sqrt \kappa  {e^{ - i{\Delta _H}t}}(\kappa  - 2ia)[ {{\varepsilon _m}\lambda {e^{i{\Delta _H}t}} + {\varepsilon _n}(\lambda  - i{\Delta _H})} ]}}{{\lambda b}} \\
&+ \frac{{8\sqrt \kappa  G{e^{i\theta }}[ {{\varepsilon _m}\lambda {e^{ - i{\Delta _H}t}} + {\varepsilon _n}(\lambda  + i{\Delta _H})} ]}}{{\lambda b}}.\label{wentaijie5}
\end{aligned}
\end{equation}
\end{small}

\begin{figure}[t]
\centerline{
\includegraphics[width=8.6cm, height=3.6cm, clip]{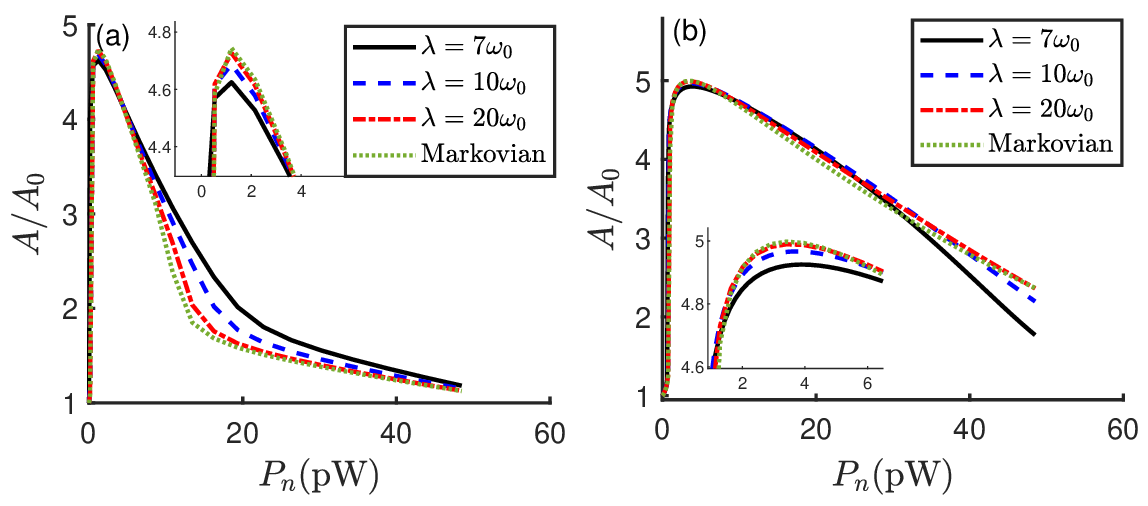}}
\caption{The influence of the environmental spectral width $\lambda$ ($\omega _g={{\omega _m} + {\omega _n}})$. The figure shows the relative response amplitude $A/A_{0}$ of the system given by Eq.~(\ref{q}) as a function of the power $P_{n}$ of the optical signal, which corresponds to the Markovian and non-Markovian environments with the different environmental spectrum width $\lambda$ and DOPA involvement ($G = 0.02 \omega_0$ and $\theta = 3\pi /2$). The frequency corresponds to (a) $\Delta_H=120 \omega$; (b) $\Delta_H=140 \omega$. Other parameters are the same as  Fig.~\ref{sgzfig2}.}\label{fmsgzfig17}
\end{figure}
\begin{figure}[t]
\centerline{
\includegraphics[width=8.6cm, height=3.6cm, clip]{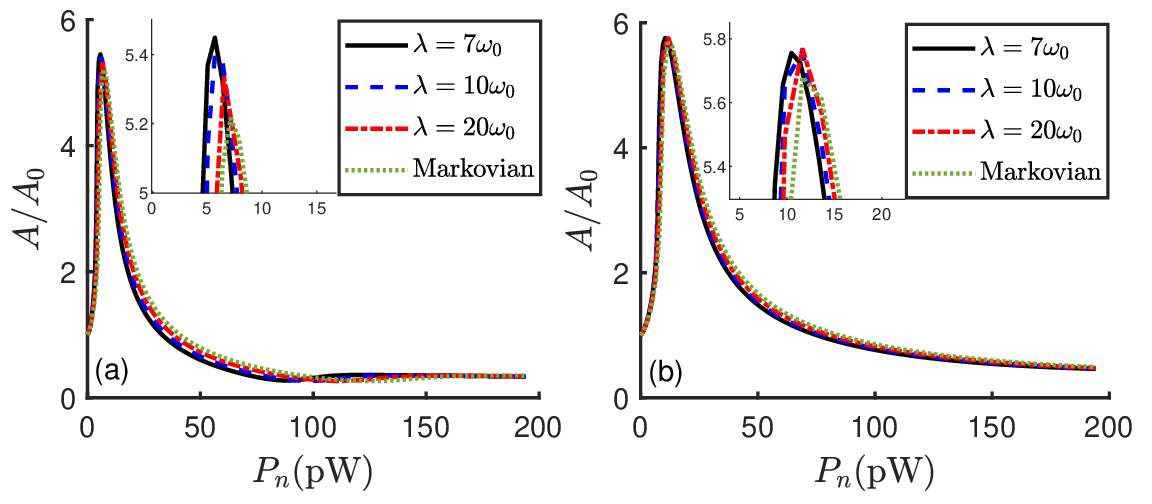}}
\caption{Impact of environmental spectral width $\lambda$ ($\omega _g=2{\omega _m}$). The relative response amplitude $A/A_{0}$ of the system determined by Eq.~(\ref{q}) respectively corresponds to the Markovian and non-Markovian environments with the different environmental spectrum width $\lambda$ and DOPA involvement ($G = 0.02 \omega_0$ and $\theta = 3\pi /2$). The frequency takes (a) $\Delta_H=80 \omega$; (b) $\Delta_H=100 \omega$. Other parameters are the same as  Fig.~\ref{sgzfig2}.}\label{fmsgzfig18}
\end{figure}

In Fig.~\ref{fmsgzfig17}(a) with $\Delta_H=120 \omega$ and the presence of the DOPA for $G = 0.02 \omega_0$ and $\theta = 3\pi /2$, we show the relative response amplitude $A/A_{0}$ of the system versus the amplitude of the optical signal $P_{n}$ with the different spectral width $\lambda$ of environment. To clearly present the influences of the non-Markovian effect on weak force amplification, Fig.~\ref{fmsgzfig17}(a) contains enlarged figure of the peak area. For a given spectral width of the environment, decreasing from ${\lambda} = 20{\omega _0}$ to $7{\omega _0}$, we find from the figure that the relative response amplitude $A/A_{0}$ gradually decreases in the non-Markovian environment. From Fig.~\ref{fmsgzfig17}(b), when the frequency $\Delta_H=140 \omega$, the relative response amplitude $A/A_{0}$ also decreases in the non-Markovian environment. Compared to the Markovian environment, the non-Markovian amplification effect of the weak mechanical signal by the optical signal gradually decreases. In enlarged figure, when the environmental spectral width $\lambda$ decreases from ${\lambda} = 20{\omega _0}$ to $7{\omega _0}$, the $A/A_{0}$ corresponding to the peak drops from 5 to 4.9.

\subsection{Amplification of weak low-frequency mechanical signal by the DOPA with $\omega _g=2{\omega _m}$ in non-Markovian systems}
Now, we vary the frequency of the DOPA, which is excited by a pump field with the frequency $\omega _g=2{\omega _m}$ \cite{Li100023838}. The total Hamiltonian of the system in the rotating frame at the laser frequency ${{\omega _m}}$ is given by
\begin{small}
\begin{equation}
\begin{aligned}
{{\hat H}} =& \frac{1}{2}m\omega _0^2{{\hat x}^2} + \frac{{{{\hat p}^2}}}{{2m}} +  \hbar\Delta_0 {{\hat a}^\dag }\hat a - \hbar \xi {{\hat a}^\dag }\hat a\hat x \\
&- i\hbar \sqrt \kappa  {\varepsilon _m}( {\hat a - {{\hat a}^\dag }} ) - i \hbar\sqrt \kappa  {\varepsilon _n}( {{e^{i\Delta_H t}}\hat a - {e^{ - i\Delta_H t}}{{\hat a}^\dag }} ) \\
&+ i\hbar G({{\hat a}^{\dag 2} }{e^{i\theta }} - H.c.) + {f}\cos ( {{\omega}t} )\hat x \\
&+ \hbar \sum\limits_k {{\Delta _k}\hat b_k^\dag {{\hat b}_k}}  + i\hbar \sum\limits_k {({g_k}\hat a} \hat b_k^\dag  - g_k^*{{\hat a}^\dag }{{\hat b}_k})
.\label{H6}
\end{aligned}
\end{equation}
\end{small}Based on the above Hamiltonian, the evolution of the system is governed by

\begin{figure}[t]
\centerline{
\includegraphics[width=8cm, height=6.4cm, clip]{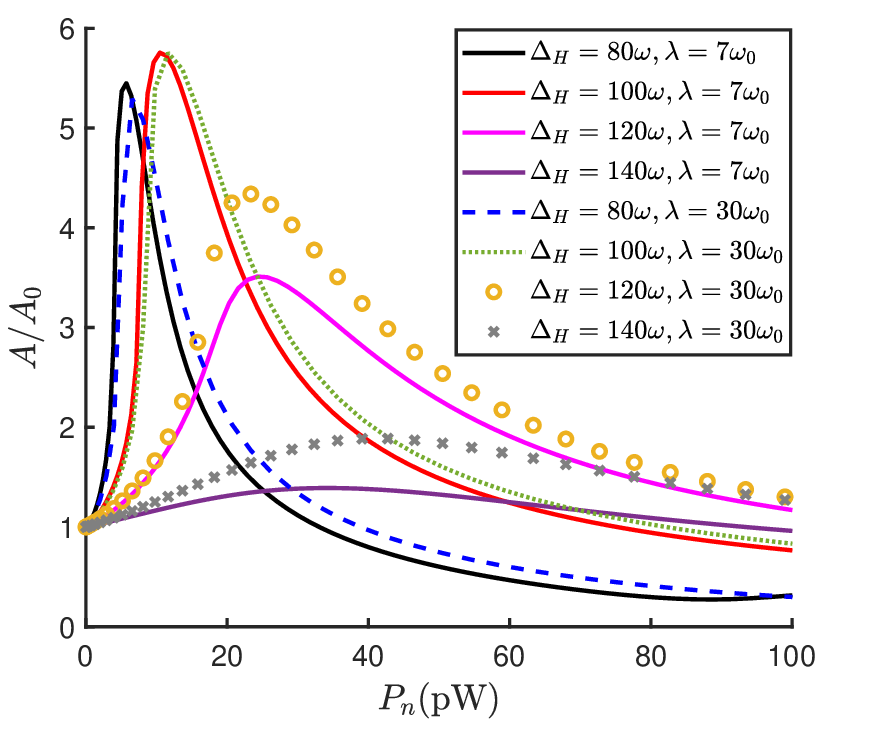}}
\caption{Different $\Delta_H$ and $\lambda$ with DOPA in non-Markovian environment ($\omega _g=2{\omega _m}$). The relative response amplitude $A/A_{0}$ of the system is calculated by Eq.~(\ref{q}) as a function of the power $P_{n}$ of the optical signal under different values of $\Delta_H$ in the non-Markovian environment and the participation of the DOPA ($G = 0.02 \omega_0$ and $\theta = 3\pi /2$). The environmental spectrum widths take $\lambda = 7{\omega _0}$ and $\lambda = 30{\omega _0}$. Other parameters are the same as  Fig.~\ref{sgzfig2}.}\label{fmsgzfig19}
\end{figure}

\vspace{-1\baselineskip}
\begin{align}
\dot \beta  =&  - i\Delta_0 \beta  + i\xi \beta x + \sqrt \kappa  {\rm{(}}{\varepsilon _m} + {\varepsilon _n}{e^{ - i\Delta_H t}}) \nonumber\\
&+ 2G{e^{i\theta }}{\beta ^*}- \sum\limits_k {g_k^*{b_k}}
,\label{Heq39}\\
\dot x =& \frac{p}{m}
,\label{Heq40}\\
\dot p =&  - {\gamma}p - m\omega _0^2x + \hbar \xi |\beta {|^2} - {f}\cos ( {{\omega}t} )
,\label{Heq41}\\
{{\dot b}_k} =&  - i{\Delta _k}{b_k} + {g_k}\beta
.\label{Heq42}
\end{align}By substituting the solution of Eq.~(\ref{Heq42}) into Eq.~(\ref{Heq39}), we get
\begin{equation}
\begin{aligned}
\dot \beta  =& - i\Delta_0 \beta  + i\xi \beta x + \sqrt \kappa  {\rm{(}}{\varepsilon _m} + {\varepsilon _n}{e^{ - i\Delta_H t}})\\
&+ 2G{e^{i\theta }}{\beta ^*} - \int_0^t {d\tau } \beta (\tau )C(t - \tau )
,\label{jifenweifenfangcheng3}
\end{aligned}
\end{equation}
and the steady-state solution with the non-Markovian effect
\begin{small}
\begin{equation}
\begin{aligned}
\beta  =& {\text{ }}\frac{{2\sqrt \kappa  {e^{ - i{\Delta _H}t}}(\kappa  - 2ia)[ {{\varepsilon _m}\lambda {e^{i{\Delta _H}t}} + {\varepsilon _n}(\lambda  - i{\Delta _H})} ]}}{{\lambda b}} \\
&+ \frac{{8\sqrt \kappa  G{e^{i\theta }}[ { {{\varepsilon _m}\lambda  + {\varepsilon _n}(\lambda  + i{\Delta _H}){e^{i{\Delta _H}t}}}} ]}}{{\lambda b}}
.\label{wentaijie6}
\end{aligned}
\end{equation}
\end{small}

\begin{figure}[t]
\centerline{
\includegraphics[width=8cm, height=6.4cm, clip]{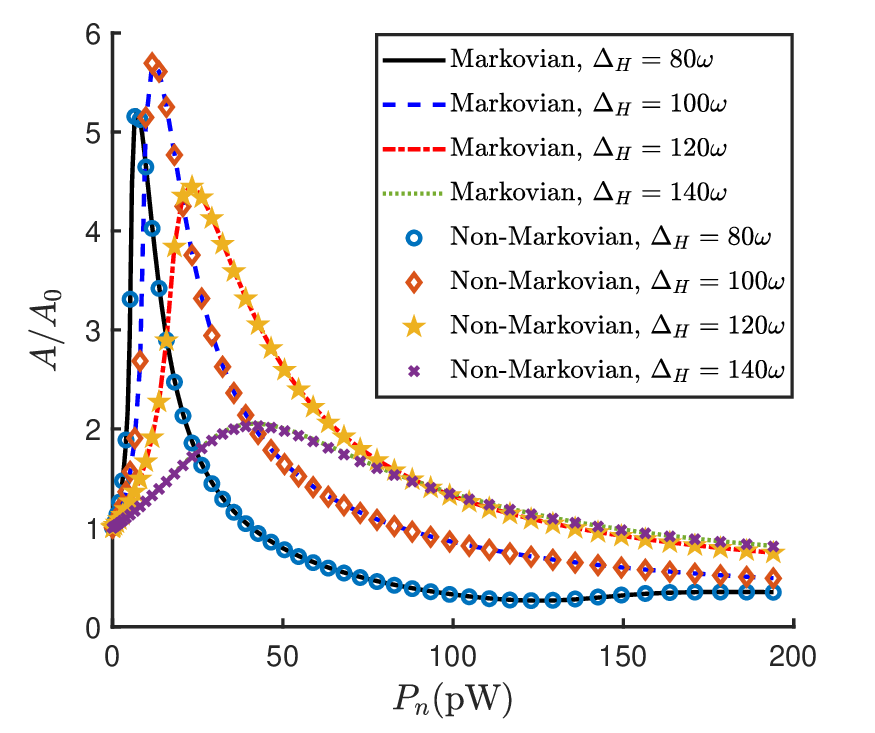}}
\caption{The non-Markovian limit and Markovian cases ($\omega _g=2{\omega _m}$). The relative response amplitude $A/A_{0}$ of the system is solved by Eq.~(\ref{q}) as a function of the power $P_{n}$ of the optical signal under different values of $\Delta_H$, where we take $G = 0.02 \omega_0$ and $\theta = 3\pi /2$. This figure shows the consistency of the relative response amplitude between non-Markovian limit with $\lambda = 200{\omega _0}$ and Markovian approximation. Other parameters are the same as  Fig.~\ref{sgzfig2}.}\label{fmsgzfig20}
\end{figure}

\begin{figure}[t]
\centerline{
\includegraphics[width=8.6cm, height=7.2cm, clip]{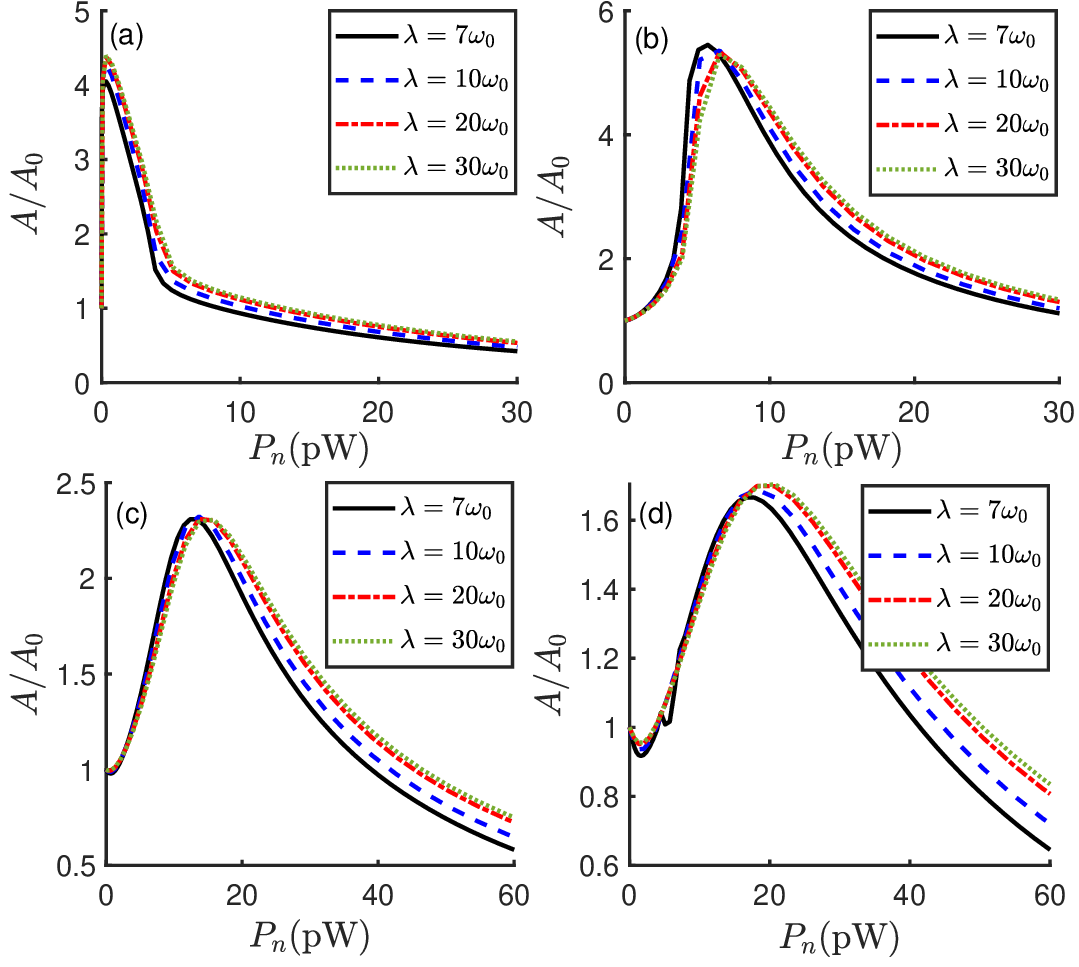}}
\caption{Effect of strength $G$ ($\omega _g=2{\omega _m}$). The relative response amplitude $A/A_{0}$ of the system given by Eq.~(\ref{q}) as a function of the power $P_{n}$ of the optical signal for different strength $G$ of the DOPA is shown in the figure, where $\theta = 3\pi /2$ and $\Delta_H= 80 \omega$. The strength $G$ respectively takes (a) $G = 0$, (b) $G = 0.02 \omega_0$, (c) $G = 0.04 \omega_0$, and (d) $G = 0.06 \omega_0$. Other parameters are the same as  Fig.~\ref{sgzfig2}.}\label{fmsgzfig21}
\end{figure}

In Fig.~\ref{fmsgzfig18}, we choose the same DOPA strength as in Fig.~\ref{fmsgzfig17} to compare two different spectral widths of environment cases ($\omega_g={{\omega _m} + {\omega _n}}$ and $\omega_g=2{\omega _m}$). In Fig.~\ref{fmsgzfig18}(a) with $\Delta_H=80 \omega$ and the presence of the DOPA for $G = 0.02 \omega_0$ and $\theta  = 3\pi /2$, we show the relative response amplitude $A/A_{0}$ of the system versus the amplitude of the optical signal $P_{n}$ with the different spectral width $\lambda$ of environment. When the spectral width of the environment decreases from ${\lambda} = 20{\omega _0}$ to $7{\omega _0}$, we find from the figure that the relative response amplitude $A/A_{0}$ gradually increases from 5.2 to 5.4 in the non-Markovian environment, which is different from the case in the frequency $\omega _g={{\omega _m} + {\omega _n}}$ in Fig.~\ref{fmsgzfig17}. To highlight the enhancing effect of the non-Markovian environment on the weak force amplification, we present the enlarged figure of the peak value. Interestingly, from Fig.~\ref{fmsgzfig18}(b), when $\Delta_H=100 \omega$, the relative response amplitude $A/A_{0}$ also exhibits a little increase from 5.67 to 5.75 (as shown in the enlarged figure) in the non-Markovian environment. Compared to the Markovian environment, the amplification effect of the optical signal on the weak mechanical signal in the non-Markovian environment significantly increases.

Figure~\ref{fmsgzfig19} shows the relative response amplitude $A/A_{0}$ of the system as a function of the power of the optical signal $P_{n}$ for different values of $\Delta_H$, where the strength and phase of the DOPA are fixed as $G = 0.02 \omega_0$ and $\theta = 3\pi /2$. We note that on the one hand, the relative response amplitude $A/A_{0}$ is very sensitive to the environmental spectrum width when the optical frequency is high. $A/A_{0}$ increases obviously with the increase of $\lambda$ at $\Delta_H=120 \omega$ and $140 \omega$. On the other hand, the environmental spectrum width has influences on the low frequency optical signal ($\Delta_H=80 \omega$ and $100 \omega$). Compared to ${\lambda} = 30{\omega _0}$, $A/A_{0}$ has a little increase with the spectral width ${\lambda} = 7{\omega _0}$ of environment, which proves that the relative response amplitude $A/A_{0}$ can be enhanced when $\Delta_H$ is low in the non-Markovian environment.

As the spectrum width further increases, the relative response amplitude $A/A_{0}$ of the system has a change. For the sake of clarity, we separately plot the non-Markovian limit and Markovian cases where the environmental spectrum width ${\lambda} = 200{\omega _0}$ for the condition that the DOPA is fixed as $G = 0.02 \omega_0$ and $\theta  = 3\pi /2$ in Fig.~\ref{fmsgzfig20}. This figure shows the consistency of the relative response amplitude $A/A_{0}$ between non-Markovian limit with ${\lambda} = 200{\omega _0}$ and Markovian approximation. It stems from the fact that the correlation function $C(t)$ approaches $\kappa \delta (t)$ when the spectral width $\lambda$ is close to infinity, which causes Eq.~(\ref{jifenweifenfangcheng3}) in the non-Markovian regime being equal to Eq.~(\ref{Heq13}) under the Markovian approximation.

The non-Markovian effect discussed here is controlled by the dimensionless ratio \(\lambda/\kappa\). When \(\lambda\gg\kappa\), the reservoir correlation function approaches a Dirac delta function and the Markovian approximation is recovered. For finite \(\lambda\) comparable to \(\kappa\), the reservoir has a finite memory time and excitation backflow can modify the weak-force amplification. In the present simulations \(\kappa=6.2\omega_0\), while \(\lambda\) is varied from \(7\omega_0\) to \(200\omega_0\), covering both the non-Markovian regime and the wide-band Markovian limit. The specific zipper cavity analyzed in Ref.~\cite{Eichenfield459550}, with \(\Gamma/2\pi\sim6~\mathrm{GHz}\), is not the target linewidth for the non-Markovian amplification considered here. The present proposal instead targets a high-\(Q\) optomechanical platform with strong dispersive coupling and an engineered Lorentzian reservoir.

Figures~\ref{fmsgzfig18}, \ref{fmsgzfig19} and \ref{fmsgzfig20} present the influences of non-Markovian effects on the relative response amplitude with the DOPA for $G = 0.02 \omega_0$ and $\theta  = 3\pi /2$. In Fig.~\ref{fmsgzfig21}, we show the variation of $A/A_{0}$ in the strength of the DOPA. As expected, when the strength $G$ of the DOPA increases from $0$ to $0.06 \omega_0$, the relative response amplitude $A/A_{0}$ tends to increase and then decrease, which is similar to Fig.~\ref{sgzfig12}. Moreover, the strength $G$ is pronounced for $A/A_{0}$ when $G$ takes $0.02 \omega_0$ and $0.04 \omega_0$. To be specific, decreasing from ${\lambda} = 30{\omega _0}$ to $7{\omega _0}$, we find from the figure that the relative response amplitude $A/A_{0}$ gradually increases in the non-Markovian environment with $G = 0.02 \omega_0$ and $0.04 \omega_0$. But the relative response amplitude $A/A_{0}$ gradually decreases in the non-Markovian environment with $G = 0$ and $0.06 \omega_0$.

\section{Experimental Implementation}
In this section, we discuss the feasibility of the experimental implementation in the cavity-optomechanical systems. The proposed setup should be regarded as a high-\(Q\) cavity-optomechanical platform with a strong dispersive coupling scale. The baseline optomechanical VR model and parameter values are consistent with the earlier study in Ref.~\cite{Shan109023511}. The coupling coefficient is further motivated by photonic-crystal optomechanics, while the linewidth used in the simulations corresponds to a high-\(Q\) optical regime rather than to the specific low-\(Q\) zipper-cavity linewidth of Ref.~\cite{Eichenfield459550}. The DOPA is modeled as an effective intracavity parametric interaction acting on the optical mode, and the Lorentzian non-Markovian reservoir can be represented through a pseudomode (see Appendix~\ref{APP.3}) or structured-reservoir construction. Therefore, the parameter set is not a direct copy of one particular experiment, but a consistent high-\(Q\) optomechanical regime in which the different ingredients of the proposed amplification mechanism can be combined. For the model under study, we mainly focus on the following three points: (i) two-photon interaction, (ii) probing the non-Markovian effect, and (iii) Lorentzian spectral density.

(i) In the parametric amplification system, we can realize the coupling between the optical cavity and the DOPA \cite{Nation841,Leghtas347853,Clerk821155,Adiyatullin813292017,Boyd2008}. This system contains two photons (representing the signal $\omega_p$ and the idler $\omega_q$ respectively), which originate from the conversion of the pump photon $\omega_g$ and satisfy the frequency relation $\omega_g=\omega_p+\omega_q$. When the cavity is driven by a classical pump (such as a laser or a microwave generator), this process is what we know as parametric down-conversion in a ${\chi ^{(2)}}$ nonlinear dielectric medium \cite{Boyd2008}, which will not be significantly damped due to the photon loss in the down-conversion process. In this case, the Hamiltonian of the system can be written as \cite{Nation841}: ${\hat H_{OPA}} = \hbar (g\hat a_p^\dag \hat a_q^\dag  + {g^*}{\hat a_p}{\hat a_q})$, where $g$ is the pump coupling coefficient. In the degenerate case where the signal mode and the idler mode coincide, i.e., ${{\hat a}_p} = {{\hat a}_q} = \hat a$, the corresponding Hamiltonian is

\vspace{-1\baselineskip}
\begin{equation}
\begin{aligned}
{\hat H_{DOPA}} = \hbar (g{\hat a^{\dag 2}} + {g^*}{\hat a^2})
,\label{GG}
\end{aligned}
\end{equation}
which indicates that the pump drives the cavity mode at twice the resonant frequency of the cavity mode.

(ii) The non-Markovian system represents a cavity interacting with an environment that retains memory effects, where all environmental feedback influences the quantum system's dynamics. Experimental characterization of such systems involves analyzing the spectral density of a condensed-matter thermal environment through the non-Markovian response of an optomechanical resonator interacting with it. This has been achieved experimentally by monitoring the optical output of a micro-optomechanical setup by applying system identification techniques \cite{groblacher676062015}. The experimental setup features a ${\rm{S}}{{\rm{i}}_{\rm{3}}}{{\rm{N}}_{\rm{4}}}$ membrane with a highly reflective central mirror segment serving as the mechanical oscillator. Additional experimental studies on non-Markovian dynamics are documented in Refs. \cite{Xu201082,hoeppe1080436032012,madsen1062336012011,liu79312011,xu172010,orieux585752015}. Through these experimental approaches, the specific form of the environmental spectral density can be determined.

(iii) To clarify how the decayed information can flow back to the system for the Lorentzian spectral density, we make use of the so-called pseudomode theory developed in Refs. \cite{Jack630438032001,Barnett1997,Garraway552290,garraway5546361997,man900621042014,man2357632015,mazzola800121042009,pleasance960621052017}. Based on the pseudomode theory, pseudomodes of an environment are auxiliary variables, which are introduced according to the position of the poles in the environment's spectral function. The interaction between the cavity and environment can be considered to be coherent interaction between the cavity mode and pseudomodes, which are surrounded by the external Markovian environments known as the pseudomodes' environments. By adding pseudomodes to the cavity as a hybrid quantum system, we derive the steady-state solution with Lorentzian spectral density. Also, for the system's quantum noise, the Lorentzian spectral density corresponds to the cavity mode undergoing a Gaussian Ornstein-Uhlenbeck process when the environment is in the vacuum state \cite{uhlenbeck368231930,gillespie5420841996,jing1052404032010}.

Finally, we suggest that the weak-force amplification in non-Markovian systems may be realized by putting an optical cavity into a ${\chi ^{(2)}}$ nonlinearity medium (or degenerate optical parametric amplifier) and coupling with the Lorentz-type spectral density environment (which is realized by pseudomode method or Gaussian Ornstein-Uhlenbeck process), where the optical cavity is driven by the external laser fields. In this sense, the prediction presented here is observable within the current experimental technology.

\section{Conclusions and Discussions}
In summary, we have explored weak-force amplification in cavity-optomechanical systems by applying DOPA within the Markovian framework. The effect of a DOPA under various pumping frequencies on the relative response amplitude was systematically analyzed. The relative response amplitude is enhanced in the presence of the DOPA, which can be tuned by the pumping frequency, the phase, and strength of the DOPA. With a weak low-frequency mechanical signal amplified by a high-frequency mechanical signal, the higher strength of the DOPA is, the higher values of $A/A_{0}$ are. When the DOPA is excited by a pump field with the frequency $\omega_g={{\omega _m} + {\omega _n}}$, similar to that of the mechanical response, the relative response amplitude $A/A_{0}$ increases as the strength of the DOPA rises. When the DOPA is excited by a pump field with the frequency $\omega_g=2{\omega _m}$, the amplification effect exhibits a trend of initially increasing and then decreasing as the strength of the DOPA increases. Moreover, we extended the study of the weak-force amplification from the Markovian environment to the non-Markovian one, which consists of a collection of infinite oscillators (bosonic photonic modes). We found the relative response amplitude $A/A_{0}$ exhibits a conversion from the non-Markovian regime to Markovian regime by controlling environmental spectral width. In the non-Markovian environment, $A/A_{0}$ decreases when a weak mechanical signal is amplified by a high-frequency mechanical signal. When the weak mechanical signal is amplified by an optical signal, $A/A_{0}$ shows variations under different pumping frequencies, exhibiting either an increase or a decrease.

These results indicate the benefits of using cavity-optomechanical systems and enhance understanding of signal amplification in the non-Markovian environment, which provides potential applications in weak signal detection, optical signal processing, and quantum information science. Expansions of the above DOPA and non-Markovian amplification effects to other nonlinear physical models, such as (1) second-order nonlinear mediums $p{\hat m^2}{\hat n^\dag } + {p^*}\hat n{\hat m^{\dag 2}}$ \cite{Majumdar235319,Shen023849,Zhou023838}, (2) third-order nonlinear materials $G{{\hat m}^{\dag 2}}{{\hat m}^2}$ \cite{zz3,zz4,chen0336032024}, (3)  quadratic cavity-optomechanical coupling systems $Q{{\hat m}^\dag }\hat m(\hat n + {{\hat n}^\dag })^2$ \cite{Aspelmeyer861391,Thompson45272,jack630438032001zs,jack630438032001zzz,jack630438032001z1s,tang111063513,jack630438032001zs3}, and (4) anisotropic non-rotating-wave interaction systems $\sum\nolimits_k {{g_k}( {{{\hat m}_k}{{\hat n}^\dag } + \hat m_k^\dag \hat n} ) + {J_k}( {{{\hat m}_k}\hat n + \hat m_k^\dag {{\hat n}^\dag }} )}$ \cite{xie0210462014,Chen752024,Chen0437082021,Rodriguez0466082008,Ai0421162010,Zheng5592004,Lu0543022007}, merit future explorations.

The cavity optomechanical VR amplification mechanism manipulated by the DOPA proposed in this paper opens up a new potential application field for quantum sensing \cite{Chowdhury112400,Gavartin7509,Krause6768}, quantum communication \cite{Stannigel105220501}, and quantum information processing \cite{Lakhfif518129678,Yanagimoto40103332023}, enabling us to better understand the relationships between DOPA parametric manipulation and weak force amplification, as well as the non-Markovian memory effect \cite{liu79312011,breuer880210022016,groblacher676062015,bylicka457202014} and the stability of signal amplification.

\section*{ACKNOWLEDGMENTS}
We would like to thank Dr.$ $ Bi-Xuan Fan and Dr.$ $ Hai-Jun Xing for valuable and constructive discussions. This work was supported by Science and Technology Development Plan Project of Jilin Province (Grant No.~20250102007JC), National Natural Science Foundation of China (NSFC) under Grant No. 12274064 and No. 12374333, and Hakubi Projects of RIKEN.

\section*{DATA AVAILABILITY}
The data that support the findings of this article are not publicly available. The data are available from the authors upon reasonable request.

\appendix
\section{The effective potential for weak-force amplification in cavity-optomechanical systems}\label{APP.1}
Equation~(\ref{x1}) can be decomposed into a slow motion $X(t)$ and a fast motion $\varphi(t,\tau)$, where the evolution equation of the slow motion $X(t)$ can be written as
\begin{small}
\begin{align}
&\ddot X + \gamma \dot X =  - \omega _0^2X - \frac{{f\cos ( {\omega t} )}}{m} \nonumber\\&+ \frac{{c[{e^{ - i\theta }}(\kappa  - 2ia') + 4G][{e^{i\theta }}(\kappa  + 2ia') + 4G]}}{{m{b^2}}}\nonumber\\& + \frac{{d(1152{G^2}{a'^2} - 32{G^2}{\kappa ^2} + 8{a'^2}{\kappa ^2} + 48{a'^4} + 768{G^4} - {\kappa ^4})}}{{m{b'^4}}}\nonumber\\& + \frac{{8Gi{e^{ - i\theta }}d[6{\kappa ^2}a' + i{\kappa ^3} - 24a'(4{G^2} + {a'^2}) - 4i\kappa (4{G^2} + 5{a'^2})]}}{{m{b'^4}}} \nonumber\\& - \frac{{8Gi{e^{i\theta }}d[6{\kappa ^2}a' - i{\kappa ^3} - 24a(4{G^2} + {a'^2}) + 4i\kappa (4{G^2} + 5{a'^2})]}}{{m{b'^4}}}
,\label{X1}
\end{align}
\end{small}where $a' = {\Delta _0} - \xi X$, $b' = 4{a'^2} - 16{G^2} + {\kappa ^2}$, and $d = 16\varepsilon _m^2{\xi ^3} \hbar\kappa \langle {{\varphi ^2}} \rangle$. Moreover, the equation for the fast dynamics of the system can be obtained by subtracting Eq.~(\ref{X1}) from the original equation in Eq.~(\ref{x1markovian}) as follow
\begin{equation}
\begin{aligned}
\ddot \varphi  + {\gamma}\dot \varphi  \approx \frac{{ - {F}\cos ( {{\Omega}t} )}}{m}
.\label{fai2}
\end{aligned}
\end{equation}

Assuming $\Omega$ is a large frequency, we can neglect the terms containing $\varphi$ and its higher-order terms based on the fact that $\ddot{\varphi} \gg \dot{\varphi} \gg \varphi, \varphi^{2}, \varphi^{3}$. From Eq.~(\ref{fai2}), we find
\begin{equation}
\begin{aligned}
\left\langle\varphi^{2}\right\rangle=\frac{F^{2} / m^{2}}{2(\Omega^{4}+\gamma^{2} \Omega^{2})}
.\label{psi2average}
\end{aligned}
\end{equation}
The effective potential ${V_{{\text{eff}}}}(X)$ corresponding to the slow motion of the system can be derived by $m( {\ddot X \!+\! \gamma \dot X} ) = {{ - \partial V(X)} \mathord{\left/
 {\vphantom {{ \! - \partial V(X)} {\partial X}}} \right.
 \kern-\nulldelimiterspace} {\partial X}}$ from Eq.~(\ref{X1})
\begin{align}
&{V_{{\text{eff}}}}(X) = \frac{{m\omega _0^2{X^2}}}{2} + f\cos ( {\omega t} )X \nonumber\\&- \frac{{2\hbar\varepsilon _m^2{\kappa ^2} [ {2G( {{e^{i\theta }}+e^{ - i\theta }}) + \kappa } ] \vartheta }}{{{{( {{\kappa ^2} - 16{G^2}} )}^{3/2}}}}\nonumber \\
&+ \frac{{4G{\kappa ^4}U[2a'({e^{i\theta }}+e^{ - i\theta }) - i\kappa ( {e^{i\theta }} - {e^{ - i\theta }})]}}{b'} \nonumber\\&- \frac{{64{G^2}\kappa U[a'(16{G^2} - {\kappa ^2}) + 16i{G^3}({e^{i\theta }} - {e^{ - i\theta }})]}}{b'} \nonumber \\
&+ \frac{{128U{G^3}{\kappa ^2}[i\kappa ( {e^{i\theta }} - {e^{ - i\theta }}) - a'({e^{i\theta }}+e^{ - i\theta })]}}{b'} \label{ueff1}\\&+ \frac{{16{\kappa}GW[i\kappa ( {e^{i\theta }}- {e^{ - i\theta }}) - 2a'({e^{i\theta }}+e^{ - i\theta }) ]}}{{{b'^2}}}\nonumber\\
&- \frac{{256{G^2} W\left[ {a' + iG\left( { {e^{i\theta }} - {e^{ - i\theta }} } \right)} \right]}}{{{b'^2}}} \nonumber\\&- \frac{{8 a'W\left[ {6G\kappa \left( {{e^{i\theta }}+e^{ - i\theta }} \right) + 32{G^2} + {\kappa ^2}} \right]}}{{\left( {{\kappa ^2} - 16{G^2}} \right)b}}\nonumber \\
&+ \frac{{4 W\left[ {{32{G^2} + {\kappa ^2}} + 6G\kappa ({e^{i\theta }} +  {e^{ - i\theta }})} \right] \vartheta }}{{{{\left( {{\kappa ^2} - 16{G^2}} \right)}^{3/2}}}},\nonumber
\end{align}
where $\vartheta  = \arctan ( { - 2a'{\text{ }}/\sqrt {{\kappa ^2} - 16{G^2}} } )$, $U = \varepsilon _m^2\hbar /{({\kappa ^2} - 16{G^2})^2}$, and $W = \kappa \varepsilon _m^2{\xi ^2}\hbar \left\langle {{\varphi ^2}} \right\rangle /({\kappa ^2} - 16{G^2})$. Here, ${V_0}(X)$ is the potential function in the absence of the high-frequency mechanical force \cite{Xie98052202}, which corresponds to
\begin{align}
&{V_0}(X) = \frac{{m\omega _0^2{X^2}}}{2} + f\cos \left( {\omega t} \right)X \nonumber \\
&- \frac{{2\hbar\varepsilon _m^2{\kappa ^2} \left[ {2G\left( {{e^{i\theta }}+{e^{ - i\theta }}} \right) + \kappa} \right] \vartheta}}{{{{\left( {{\kappa ^2} - 16{G^2}} \right)}^{3/2}}}} \nonumber\\& + \frac{{4G{\kappa ^4}U[2a'({e^{i\theta }}+{e^{ - i\theta }} ) - i\kappa ( {e^{i\theta }} - {e^{ - i\theta }})]}}{b'} \label{u0}\\&- \frac{{64{G^2}\kappa U[a'(16{G^2} - {\kappa ^2}) + 16i{G^3}({e^{i\theta }} - {e^{ - i\theta }})]}}{b'} \nonumber \\
&+ \frac{{128U{G^3}{\kappa ^2}[i\kappa ( {e^{i\theta }} - {e^{ - i\theta }}) - a'({{e^{i\theta }}+e^{ - i\theta }})]}}{b'}.\nonumber
\end{align}
Comparing Eqs.~(\ref{ueff1}) with (\ref{u0}), there is an additional term (the last four terms), which represents the contribution of the high-frequency mechanical force to the potential function. Essentially, the high-frequency signals can effectively alter the nonlinear characteristics of the system, which offers the potential to decrease the potential barrier and amplify the system response to the weaker signal.
When driven by high-frequency mechanical signals, the non-Markovian and Markovian environments have the same steady state solution. The effective potential function $V_{{\text{eff}}}(X)$ of the non-Markovian environment can be obtained by substituting Eq.~(\ref{wentaijie4}) into Eq.~(\ref{Heq25}), which is the same as in the Markovian environment, as shown in the above derivation given by Eqs.~(\ref{X1})-~(\ref{u0}).

\section{derivation of Eq.~(\ref{ft})
}\label{APP.2}
Substituting Eq.~(\ref{bk1}) into Eq.~(\ref{Heq23}) gives
\begin{small}
\begin{equation}
\begin{aligned}
\dot \beta =& - i{\Delta _0}\beta  + i\xi \beta x + \sqrt \kappa  {\varepsilon _m} + 2G{e^{i\theta }}{\beta ^*} \\
&- \sum\limits_k {g_k^*[{b_k}(0){e^{ - i{\Delta _k}t}}+ {g_k}\int_0^t {d\tau } \beta (\tau ){e^{ - i{\Delta _k}(t - \tau )}}]} \\
=& - i{\Delta _0}\beta  + i\xi \beta x + \sqrt \kappa  {\varepsilon _m} + 2G{e^{i\theta }}{\beta ^*} \\
&- \sum\limits_k {g_k^*} {b_k}(0){e^{ - i{\Delta _k}t}} \\
&+ \int_0^t {d\tau } \beta (\tau )\sum\limits_k {{{\left| {{g_k}} \right|}^2}} {e^{ - i{\Delta _k}(t - \tau )}} \\
\equiv &  - i\Delta_0 \beta  + i\xi \beta x + \sqrt \kappa  {\varepsilon _m} + 2G{e^{i\theta }}{\beta ^*} + {{Z}}(t)\\
& - \int_0^t {d\tau } \beta (\tau )C(t - \tau )
,\label{jifenweifenfangchengsuanfub}
\end{aligned}
\end{equation}
\end{small}where ${{Z} }(t) =  - \sum\nolimits_k {g_k^*{{ b}_k}(0){e^{ - i{\Delta _k}t}}}$, while the correlation function is given by
\begin{equation}
\begin{aligned}
C(t) = \sum_k {{{\left| {{g_k}} \right|}^2}{e^{ - i{\Delta _k}t}}}.
\label{ftb}
\end{aligned}
\end{equation}
Introducing the spectral density $J(\omega')$ of the environment \cite{breuer2002,breuer880210022016,breuer1032104012009,Vacchini042103}
\begin{equation}
\begin{aligned}
J(\omega') = \sum_k |g_k|^2 \delta(\omega' - \omega_k),
\label{Jb}
\end{aligned}
\end{equation}
$C(t)$ in Eq.~(\ref{ftb}) can be rewritten as
\begin{equation}
\begin{aligned}
C(t) =& \sum_k |g_k|^2 e^{-i(\omega_k - \omega_m)t} \\
&= \int d\omega' \, [ \sum_k |g_k|^2 \delta(\omega' - \omega_k) ] e^{-i(\omega' - \omega_m)t}\\
&= \int J(\omega') e^{-i(\omega' - \omega_m)t}  d\omega',
\label{ctuidao}
\end{aligned}
\end{equation}
which corresponds to Eq.~(\ref{ft}).

\section{The impulse response function $h(t)$ and correlation function $C(t)$ can be realized through the pseudomode theory
}\label{APP.3}
Non-Markovian effects with a Lorentzian spectral density can be exactly mapped to a system where the cavity mode couples to a single pseudomode, which in turn dissipates into a Markovian environment \cite{Jack630438032001,Barnett1997,Garraway552290,garraway5546361997,man900621042014,man2357632015,mazzola800121042009,pleasance960621052017}.
The pseudomode acts as a mediator: when its decay rate is small, excitations can coherently exchange between the cavity and the pseudomode, leading to backflow and non-Markovian behavior. When the pseudomode decay rate is large, the dynamics becomes Markovian. Now we go into the details. We consider cavity mode (eigenfrequency ${\omega _a}$) interacting with a pseudomode (eigenfrequency $\omega _u$). The total Hamiltonian in Eq.~(\ref{H1}) becomes
\begin{align}
{{\hat H}_{S}}=& \frac{1}{2}m\omega _0^2{{\hat x}^2} + \frac{{{{\hat p}^2}}}{{2m}} +  \hbar\Delta_0 {{\hat a}^\dag }\hat a - \hbar \xi {{\hat a}^\dag }\hat a\hat x \nonumber \\
&- i \hbar\sqrt \kappa  {\varepsilon _m}( {\hat a - {{\hat a}^\dag }} ) + i\hbar G({{\hat a}^{\dag 2} }{e^{i\theta }} - H.c.) \nonumber \\
&+ {f}\cos ( {{\omega}t} )\hat x + {F}\cos ( {{\Omega}t} )\hat x \nonumber \\
&+ \hbar {\Delta u}{{\hat u}^\dag }\hat u + \hbar {g_{au}}(\hat a{{\hat u}^\dag } + {{\hat a}^\dag }\hat u),
\label{xuanzhuanqianHS}
\end{align}
where $\hat u$ denotes the pseudomode annihilation operator satisfying bosonic commutation relations $[\hat u,  {\hat u}^\dag ] = 1$. $\Delta u = {\omega _u} - {\omega _m}$ denotes the detuning between the pseudomode and driving field, which is set as zero (i.e., $\Delta u = 0$). The final term in Eq.~(\ref{xuanzhuanqianHS}) corresponds to the coherent coupling between the cavity mode and the pseudomode, where ${g_{au}}$ denotes the coupling strength.
The whole Hamiltonian including the Markovian environment takes
\begin{equation}
\begin{aligned}
{{\hat H_T}} = {{\hat H}_S} + {{\hat H}_R} + {{\hat H}_I},
\label{HIHS}
\end{aligned}
\end{equation}where ${{\hat H}_I}  = i \hbar \sum\nolimits_k {{V_k}} (\hat r_k^\dag \hat u - {{\hat u}^\dag }{{\hat r}_k})$ represents the interaction Hamiltonian between the pseudomode and the Markovian environment. ${V_k} = \sqrt {\eta /2\pi }$ is the coupling strength with decay rate $\eta$. The free Hamiltonian of the environment is given by ${\hat H_R}   = \hbar \sum\nolimits_k {{(\omega _k-\omega _m)} \hat r_k^\dag {\hat r_k}}$ with the commutation relation $[{\hat r_k}, \hat r_{k'}^\dag ] = {\delta _{kk'}}$. We define the expectation values of the operators as $\beta= \langle \hat a \rangle$, $x=\langle \hat x \rangle$, $u = \langle {\hat u} \rangle$, $r_k = \langle {\hat r}_k \rangle$, $r_{\rm in} = \langle {\hat r}_{\rm in} \rangle$, and $S(t) = \langle {\hat S}(t) \rangle$. Based on Eq.~(\ref{HIHS}), the Heisenberg-Langevin equations under the Markovian approximation can be written as \cite{Weiss2008,GardinerBerlin2000}
\begin{align}
\frac{d}{{dt}}{\beta} =& - i\Delta_0 \beta + i\xi \beta x + 2G e^{i\theta}{\beta}^* \nonumber\\
&+ \sqrt{\kappa}\varepsilon_m -i g_{au} u, \label{dota}\\
\frac{d}{{dt}} u =& - i{g_{au}}\beta - \frac{\eta }{2} u - \sqrt \eta {{ r}_{\rm in}}(t).
\label{dotb}
\end{align}
Solving Eq.~(\ref{dotb}) for $u\left( t \right)$ gives $u(t) =  u(0){e^{ - \frac{\eta }{2}t}} - i{g_{au}}\int_0^t {\beta(\tau ){e^{ -  \frac{\eta }{2}(t - \tau )}}d\tau } - \sqrt \eta  \int_0^t {{{  r}_{\rm in}}(\tau ){e^{ - \frac{\eta }{2}(t - \tau )}}d\tau }$ with ${ r_{{\rm{in}}}}(t){\rm{ = }}\sum\nolimits_k {{e^{ - i{(\omega _k-\omega _m)}t}}} {r_k}/\sqrt {2\pi }$. Substituting $u(t)$ into Eq.~(\ref{dota}), we obtain
\begin{equation}
\begin{aligned}
\dot{\beta} =& -i\Delta_0 \beta + i\xi \beta x + 2G e^{i\theta} \beta^* + \sqrt{\kappa} \varepsilon_m \\
&- \int_0^t M(t-\tau) \beta(\tau) d\tau - S(t),
\label{Heisenbergequationsofmotion3}
\end{aligned}
\end{equation}
where $S(t)= i{g_{au}} u(0){e^{ -\frac{\eta }{2}t}}-i {g_{au}}\sqrt \eta  \int_0^t {{{ r}_{\rm in}}(\tau ){e^{ - \frac{\eta }{2}(t - \tau )}}d\tau }$ is the contribution for the non-Markovian composite environment (including the pseudomode and its Markovian environment). The reason for ${S}(t) = 0$ due to the pseudomode and environment initialization in the vacuum state is provided above Eq.~(\ref{ft}). $M(t) = g_{au}^2{e^{ -\frac{\eta }{2}t}}$ in Eq.~(\ref{Heisenbergequationsofmotion3}) is the corresponding correlation function.
The Lorentzian spectral density $Y(\omega )$ corresponding to the correlation function $M(t) = g_{au}^2{e^{ -  \frac{\eta }{2}t}} \equiv \int {Y(\omega )} {e^{ - i\omega t}}d\omega $ through comparing Eqs.~(\ref{jifenweifenfangchengsuanfu}) and (\ref{Heisenbergequationsofmotion3}) is equal to $
Y(\omega)=\frac{\kappa}{2\pi}\frac{\lambda^{2}}{\lambda^{2}+\omega^{2}}$,
where $\lambda  \equiv \frac{\eta}{2}$ and $\kappa \equiv \frac{4g_{au}^2}{\eta}$ give
\begin{align}
M(t-\tau) = \frac{1}{2}\kappa \lambda e^{-\lambda (t-\tau)},
\label{R29}
\end{align}which is consistent with $C(t-\tau)$ presented above Eq.~(\ref{wentaijie4}). Therefore, Eq.~(\ref{Heisenbergequationsofmotion3}) obtained by the Markovian pseudomode method is completely equivalent to Eq.~(\ref{jifenweifenfangchengsuanfu}) in the non-Markovian regime.
\begin{figure}[t]
\centerline{
\includegraphics[width=8.8cm, height=6cm, clip]{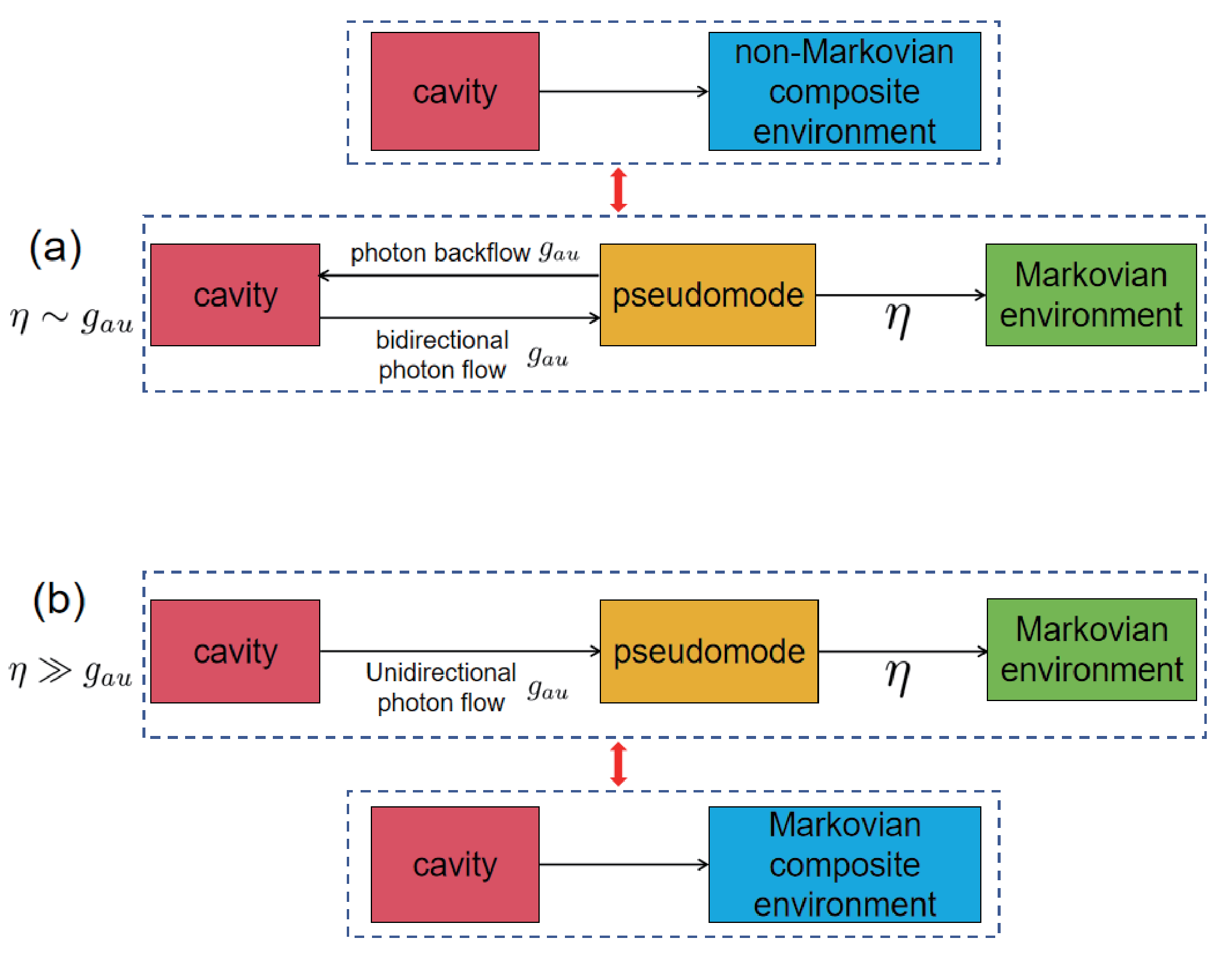}}
\caption{The figure shows the photon flow direction in the system. (a) Non-Markovian regime $(\eta \ll g_{au})$: the pseudomode decays slowly, allowing bidirectional photon exchange and backflow between the cavity and the pseudomode. (b) Markovian regime $(\eta \gg g_{au})$: the pseudomode decays rapidly, leading to unidirectional photon dissipation into the Markovian environment.}\label{liuchengtunm}
\end{figure}
The pseudomode system given by Eq.~(\ref{HIHS}) can be used to simulate and verify a non-Markovian environment with a Lorentzian spectral density. In this construction, the pseudomode decays at a rate $\eta$, and the composite environmental memory effect is characterized by the spectral width $\lambda = \eta/2$. The magnitude of $\eta$ directly determines whether the system behaves in a Markovian or non-Markovian manner. When $\eta$ is very small, the pseudomode decays slowly, allowing frequent and reversible exchange of excitations between the cavity and the pseudomode. As a result, excitations do not immediately escape into the environment. Instead, they can flow back from the pseudomode to the cavity, giving rise to strong non-Markovian effects (corresponding to non-Markovian composite environment) in Fig.~\ref{liuchengtunm}(a). This backflow does not require a photon to return from free space, which is mediated by the pseudomode, which acts as a structured memory. Conversely, when $\eta$ is large, the pseudomode decays quickly. Photons in the cavity rapidly dissipate into the Markovian environment via the pseudomode, making backflow negligible, and the system exhibits Markovian behavior (corresponding to Markovian composite environment) in Fig.~\ref{liuchengtunm}(b). Thus, by controlling decay rate $\eta$, we can smoothly tune the system from the non-Markovian to the Markovian regimes. In the non-Markovian case, the cavity first loses photons to the pseudomode, and later some excitations return, effectively realizing a Lorentzian environment. In the Markovian case, dissipation is unidirectional from environment to system. This pseudomode mechanism provides an experimentally accessible way to generate controllable non-Markovian dynamics.


\end{document}